\documentclass[acmsmall,screen]{acmart}

\usepackage{booktabs} 
\usepackage{fancyvrb}
\usepackage{listings}
\usepackage{color}
\usepackage{balance}
\usepackage{fancybox}
\usepackage{xspace}
\usepackage{subcaption}
\usepackage{caption}
\usepackage{multirow}
\usepackage{url}
\usepackage{array}
\usepackage{graphicx}
\usepackage{bbding}
\usepackage{wasysym}
\usepackage{pdflscape}
\usepackage{tabularx,colortbl}
\usepackage{dirtree}
\usepackage{rotating}

\usepackage{comment}

\usepackage{algorithm}
\usepackage[noend]{algpseudocode}
\usepackage{enumitem}

\usepackage{colortbl}
\usepackage{tabu}
\usepackage[skins]{tcolorbox}
\usepackage{cleveref}

\usepackage{balance}

\AtBeginDocument{%
  \providecommand\BibTeX{{%
    \normalfont B\kern-0.5em{\scshape i\kern-0.25em b}\kern-0.8em\TeX}}}

\definecolor{grayboxcolor}{HTML}{f2f2f2}
\newcommand{\conclusion}[1]{%
	\begin{center}\noindent\thicklines\setlength{\fboxsep}{8pt}\fcolorbox{black}{grayboxcolor}{\begin{minipage}{4.6in}\textbf{#1}\end{minipage}}\end{center}}

\definecolor{cadmiumgreen}{rgb}{0.0, 0.42, 0.24}

\def\commentsenabled{}

\ifdefined\commentsenabled
\newcommand{\diego}[1]{\textcolor{cadmiumgreen}{Diego: #1}} 
\newcommand{\rabe}[1]{\textcolor{orange}{Rabe: #1}} 
\newcommand{\abbas}[1]{\textcolor{red}{Abbas: #1}} 
\newcommand{\todo}[1]{\textcolor{red}{TODO: #1}}
\else
\newcommand{\diego}[1]{} 
\newcommand{\rabe}[1]{} 
\newcommand{\abbas}[1]{} 
\newcommand{\todo}[1]{} %
\fi

\newcommand{\rev}[1]{\textcolor{black}{#1}}

\acmJournal{TOSEM}

\begin{document}

\title{Dependency Update Strategies and Package Characteristics}

\author{Abbas Javan Jafari}
\affiliation{%
  \institution{Data-driven Analysis of Software (DAS) Lab at the Department of
  Computer Science and Software Engineering, Concordia University}
  \city{Montreal}
  \country{Canada}}
\email{a_javanj@encs.concordia.ca}

\author{Diego Elias Costa}
\affiliation{%
  \institution{LATECE Lab at the Department of Computer Science, Université du Québec à Montréal (UQAM)}
  \city{Montreal}
  \country{Canada}}
\email{costa.diego@uqam.ca}

\author{Emad Shihab}
\affiliation{%
  \institution{Data-driven Analysis of Software (DAS) Lab at the Department of
  Computer Science and Software Engineering, Concordia University}
  \city{Montreal}
  \country{Canada}}
\email{emad.shihab@concordia.ca}

\author{Rabe Abdalkareem}
\affiliation{%
  \institution{Department of Computer Science at the Faculty of Science, Omar Al-Mukhtar University}
  \city{}
  \country{Libya}}
\email{rabe.abdalkareem@omu.edu.ly}

\renewcommand{\shortauthors}{Javan Jafari et al.}

\begin{abstract}
Managing project dependencies is a key maintenance issue in software development. Developers need to choose an update strategy that allows them to receive important updates and fixes while protecting them from breaking changes. Semantic Versioning was proposed to address this dilemma but many have opted for more restrictive or permissive alternatives. This empirical study explores the association between package characteristics and the dependency update strategy selected by its dependents to understand how developers select and change their update strategies. We study over 112,000 npm packages and use \rev{19} characteristics to build a prediction model that identifies the common dependency update strategy for each package. \rev{Our model achieves a minimum improvement of 72\% over the baselines} and is much better aligned with community decisions than the npm default strategy. We investigate how different package characteristics can influence the predicted update strategy and find that dependent count, age and release status to be the highest influencing features. We complement the work with qualitative analyses of 160 packages to investigate the evolution of update strategies. While the common update strategy remains consistent for many packages, certain events such as the release of the 1.0.0 version or breaking changes influence the selected update strategy over time. 
\end{abstract}



\keywords{Dependency update strategy, Dependency management, Software ecosystems, npm}

\maketitle

\section{Introduction}
\label{sec:Introduction}
Software development is increasingly reliant on code reuse, which can be accomplished through the use of software packages. Utilizing packages to build software improves quality and productivity \cite{lim1994effects,mohagheghi2004empirical}. These packages, along with the dependencies and maintainers have formed large software ecosystems \cite{sonatype2021}. \rev{In the current landscape}, managing dependencies among packages is an emerging challenge \cite{artho2012software,bogart2016break,decan2019empirical}. The popular Node Package Manager (npm) ecosystem has experienced several dependency-related incidents. One example is the removal of the backward-incompatible release of the ``underscore'' package that generated a lot of complaints among dependents that updated to the latest version \cite{underscore2014}. Another example is the removal of the ``left-pad'' package which, at the time, majorly impacted many web services \cite{leftpad2018}. \rev{The ua-parser-js package is more a recent example of an npm package that had its maintainer account hijacked to release malicious versions of the library \citep{uaparser-github} that would steal user information such as cookies and browser passwords. The package frequently experiences 6-7 million weekly downloads and was used by many large companies such as Facebook, Apple, Amazon, Microsoft, IBM, Oracle, Mozilla, Reddit and Slack \citep{uaparser}.}

\rev{Knowing when and how to update dependencies are among the most important challenges faced by development teams \citep{tidelift2022}. The npm package manager allows for various constraints for configuring when and how each dependency will automatically update \citep{npmpkg2019}. In order to study the dynamics of dependency updates, we draw inspiration from previous literature and group the various dependency constraints into 3 update strategies: the balanced update strategy, the restrictive update strategy and the permissive update strategy \citep{decan2019package}. The specifics of each update strategy is further explained in Section~\ref{sec:Background}. Different update strategies bring about different consequences \citep{jafari2020dependency}. Opting for overly restrictive update strategies (e.g. preventing any automatic updates) will prevent timely security fixes for packages \cite{prana2021out,cox2015measuring,decan2018impact}. On the other hand, overly permissive update strategies (e.g. allowing any type of automatic updates) will increase the likelihood of breaking changes due to incompatible releases \cite{kula2018developers, decan2018evolution,jafari2020dependency}. Thus, a key issue in dependency management is choosing the right strategy for updating dependencies.}

Semantic Versioning (SemVer) has been proposed as a solution \rev{to aid dependency management by allowing maintainers to communicate the type of changes included in their new package releases and allowing developers to determine backward-compatibility based on the semantic version number of the newly released version. This provides developers with a middle-ground between keeping dependencies up to date while ensuring a backward-compatible API \cite{semver2019}. However, previous research has shown that SemVer is not always relied on in practice and it is not rare to see developers opting for alternative dependency update strategies \cite{dietrich2019dependency, chinthanet2019lag,kula2018developers,wittern2016look,cogo2019empirical}.}

\rev{Developers may adopt or modify a dependency update strategy based on their perception of a package dependency. This is visible in the dependency configuration of npm packages (package.json) where different maintainers will opt for different strategies for managing their dependencies but more importantly, a maintainer will even opt for different strategies for different dependencies in the same project \cite{jafari2020dependency}. Certain events (e.g. breaking changes) may also shift a developer's perception in regards to the previously selected update strategy \cite{cogo2019empirical}. Different dependency update strategies may be selected based on the characteristics of the target packages. Additionally, the characteristics of a package dependency may serve as indicators of the community trust on the package (e.g. age may signal maturity).} Understanding how these characteristics relate to dependency decisions among the majority of developers can serve as a guide for how one should depend on each package, as well as a means to understand what package characteristics are associated with dependency update strategies.

In this study, we investigate the relationship between npm package characteristics and the dependency update strategy opted by its dependents. We focus on npm since it currently maintains the largest number of packages in any software ecosystem \cite{libraries.io2020} and consequently, \rev{a high number of dependency relationships between packages}. Our  dataset includes 112,452 npm packages and \rev{19} characteristics derived from npm and the package repository. We use a machine learning module to investigate whether package characteristics can be used to \rev{predict} the most popular dependency update strategy for each package. Specifically, we aim to tackle the following research questions:

\textbf{RQ1:} Can package characteristics be used as indicators of dependency update strategies?

\rev{We train several machine learning models using features collected and derived from package characteristics. Our experiments reveal Random Forest as the most suitable model for our purpose. As such, we select Random Forest as the model in this paper.} We evaluate our model and compare it against two baselines (stratified random prediction and npm-recommended balanced strategy). The results show a 72\% improvement in the ROC-AUC score and 90\% improvement in the F1-score compared to the stratified baseline. We observe that package characteristics can be used as indicators of the common update strategy and they can be leveraged for predicting dependency update strategies. Additionally, we found that our model results align considerably better with community decisions than always using the balanced update strategy.

\textbf{RQ2:} Which package characteristics are the most important \rev{indicators for dependency update strategies?}

In order to help developers understand the key factors that impact dependency update strategies, we identify the most important features for the prediction model and analyze how a change in these features impacts the model's predictions. The \textit{release status} of a package, the number of \textit{dependents} and its \textit{age} (in months) are the most important indicators for the common dependency update strategy. Dependents of younger, post-1.0.0 packages with more dependents are more likely to agree on the balanced update strategy. On the other hand, dependents of pre-1.0.0 packages are more likely to opt for more permissive update strategies.

\textbf{RQ3:} How do dependency update strategies evolve with package characteristics?

In an effort to understand the prominence of evolutionary features in predicting the common update strategy, we use a mixed-method technique on a convenience sample of 160 packages to analyze the evolution of update strategies over a period of 10 years. We found that for many packages in npm, the common update strategy remains consistent throughout a package's lifecycle, but the release of the 1.0.0 version causes a visible shift in the common update strategy. Restrictive update strategies proved to experience the weakest agreement (repeatedly challenged by other strategies), with more erratic evolutionary behavior that correlate with incidents such as breaking changes.

The rest of the paper is organized as follows. Section~\ref{sec:Background} provides a background on dependency management in npm, semantic versioning and specialized packages. Section~\ref{sec:Methodology and Data} describes our data selection and feature extraction methodology. We present our results in Section~\ref{sec:Results} and highlight the study implications in Section~\ref{sec:Implications}. We review related work in Section~\ref{sec:Related Work} and discuss the threats to validity in Section~\ref{sec:Threats to Validity}. We conclude our work in Section~\ref{sec:Conclusion}.

\section{Background}
\label{sec:Background}
In this section, we present the background required to understand our work on dependency update strategies. We explain how dependencies are defined and managed in npm, explain semantic versioning, and we describe the different dependency update strategies used throughout this paper.

\subsection{Dependency management in npm}
\begin{figure}
  \centering
  \fbox{\includegraphics[width=0.70\linewidth]{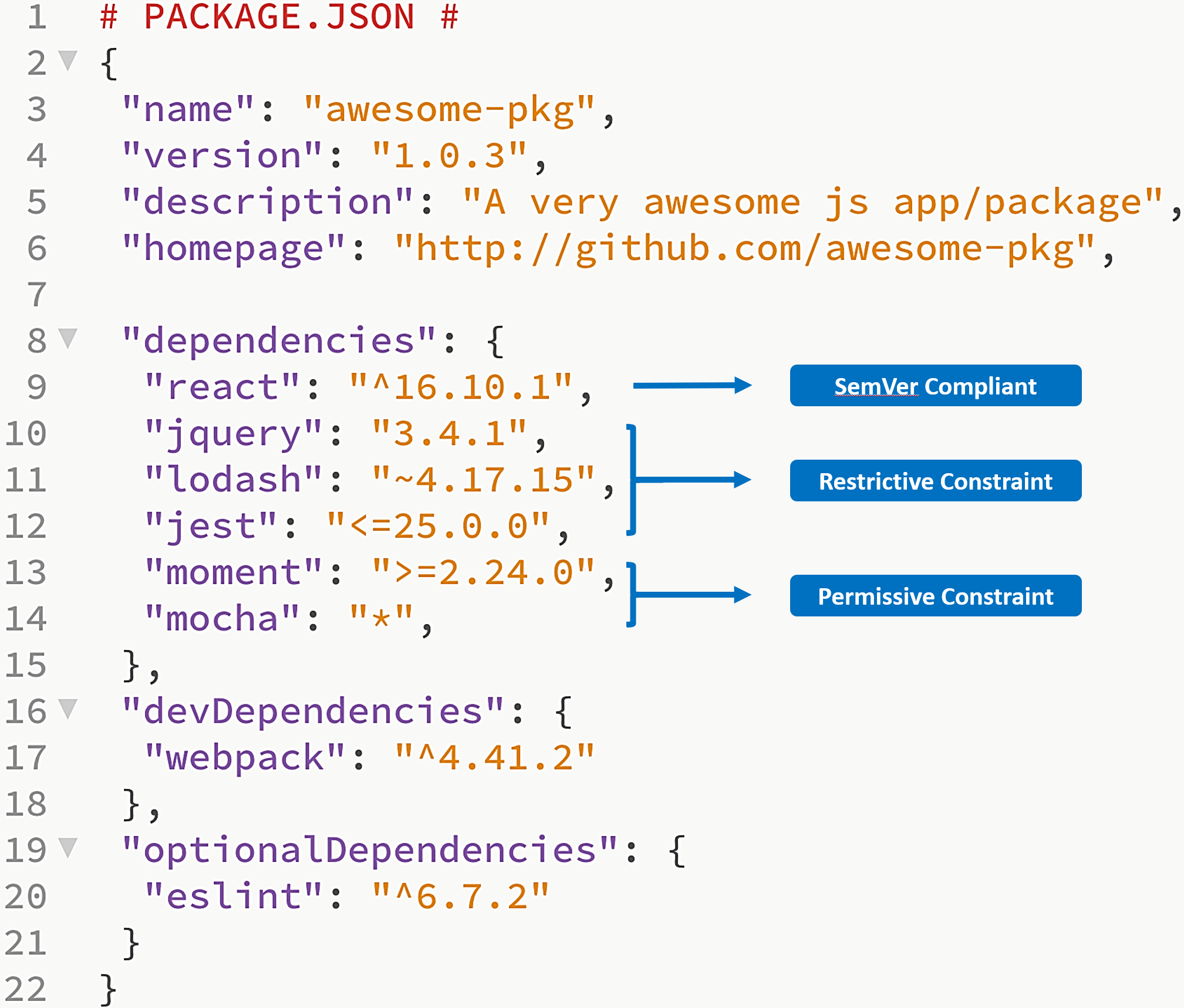}}
  \caption{Example of a package.json file showing dependency update strategies}
  \label{fig:package-json}
\end{figure}

Packages in the npm ecosystem use the package.json file to specify package metadata and the different types of dependencies \cite{npmpkg2019}. Figure~\ref{fig:package-json} depicts an example package.json file along with the three dependency update strategies referenced throughout this paper. This file uses different sections for runtime, development and optional dependencies. When a package is installed, npm will fetch and install all runtime dependencies. This is also performed for transitive dependencies (dependencies of dependencies) until the full dependency tree is installed. Upon using the \textit{npm install} command, the package manager also creates a package-lock.json which includes the installed versions of all dependencies at the time. This helps future installations of a package to remain consistent.

Our work strictly focuses on runtime dependencies since they are the dependencies required for the package to function correctly. A missing or unused package in runtime dependencies is considered bad practice as it may create runtime errors or cause extraneous installations \cite{jafari2020dependency}. Development dependencies are used for development and testing purposes. They are not required for users of the package and they are sometimes incomplete. The npm package manager will try to fetch optional dependencies, but failure to do so will not raise an error since they are also unnecessary for the package to function correctly. 

\subsection{Semantic Versioning}
Semantic Versioning (SemVer) is the \rev{de facto} versioning standard for npm \cite{npmsemver}, as well as many other software ecosystems (e.g. the PyPI ecosystem for Python). Tom Preston-Warner, the co-founder of the GitHub platform, first introduced SemVer in 2011. SemVer 2.0 was released in 2013 and it is the version used in this paper. SemVer addresses the dependency update issue by allowing package maintainers to communicate what type of changes are included in a new release. SemVer introduces a multi-part versioning scheme in the form of \textbf{major.minor.patch[-tag]}. If a newly released version contains backward incompatible feature updates, the maintainer will increase the major version number. If it includes a backward compatible feature update, they will increase the minor version number. If the new release only contains bug or security fixes, the maintainer will increase the patch version number. The optional tag is used for specifying build metadata and pre-release or post-release numbers.

Developers can use this versioning convention, along with the dependency notations in npm, to specify the degree of freedom granted to the package manager in fetching new versions of a dependency. In order to be compliant with SemVer (and assuming developers want to receive updates while avoiding breaking changes), developers should accept automatic updates for new minor and patch version for all post-1.0.0 releases. \rev{We use the term ``balanced'' to refer to such update strategies in this paper.}  The common dependency notations in npm are as follows:

\begin{itemize}
    \item The caret (\textsuperscript{$\wedge$}) notation is used to accept only minor and patch updates for post-1.0.0 versions. For example, \textsuperscript{$\wedge$}2.3.4 is equivalent to [2.3.4-3.0.0).
    \item The tilde ($\sim$) notation is used to accept only patch updates (when a minor version is specified). For example, ($\sim$)2.3.4 is equivalent to [2.3.4-2.4.0).
    \item The star {(*)} wildcard will give npm complete freedom to install any new version of a dependency.
    \item Specifying a specific version will limit npm to only install that particular version.
  \end{itemize}

\subsection{Specialized packages}

In order to identify the ``common'' dependency update strategy for a particular package, we rely on the ``wisdom of the crowds'' principle \cite{decan2019package}. This means that a dependency update strategy is deemed the common strategy if the majority of its dependents are using the same strategy. \rev{A package is deemed specialized toward an update strategy if the majority of its dependents agree on that particular update strategy.} In this paper, we calculate the proportion of each of the 3 dependency update strategies and use 50\% as the threshold to define specialized packages. If more than 50\% of the dependents are not using a common update strategy, a package is deemed unspecialized and we can not use package characteristics to analyze dependency update strategies for that package. Section 3.1 explains the rationale for the selected threshold. By drawing inspiration from the work of Decan and Mens \cite{decan2019package}, a package is considered specialized if more than 50\% of its dependents agree on one of the following update strategies:
\begin{itemize}
  \item \textbf{Balanced:} The update strategy is considered balanced if it allows for new updates but keeps us safe from breaking changes (with the assumption that SemVer is correctly followed by the target package). \rev{In specific terms, a post-1.0.0 constraint that allows automatic updates to new minor and patch versions is considered balanced. This can be accomplished by using the caret notation in npm (e.g. ``\textsuperscript{$\wedge$}1.2.3'') but can also be expressed in other ways such as ``1.x.x''. A pre-1.0.0 constraint is considered balanced if it does not allow any updates (pinned). This is due to the fact that SemVer considers these versions to have an unstable API \citep{semver2019}.}
  \item \textbf{Restrictive:} The update strategy is considered restrictive if it is more restrictive than the balanced update strategy. \rev{In specific terms, a post-1.0.0 constraint that only allows automatic updates to new patch releases or no automatic updates at all is considered restrictive. This can be accomplished through the use of the tilde notation in npm (e.g. ``$\sim$1.2.3'') but can also be expressed in other ways such as ``1.2.x'' or ``1.2.3''. Pre-1.0.0 constraints can not be restrictive since pre-1.0.0 releases have an unstable API and any freedom in updates is considered permissive.}
  \item \textbf{Permissive:} The update strategy is considered permissive if it is more permissive than balanced update strategy. \rev{In specific terms, a post-1.0.0 constraint that allows automatic updates to all new versions (including major versions) is considered permissive. This can be accomplished through the use of wildcards (e.g. ``*'') but can also be expressed in other ways such as ``latest'' or ``$>$=1.2.3''. A pre-1.0.0 constraint that allows any automatic updates is considered permissive.}
\end{itemize}

\section{Data and Methodology}
\label{sec:Methodology and Data}
We use the latest version of the libraries.io dataset available at the time of collection, containing package dependencies from January 2020\footnote{At the time of this study, no other dataset has been published since 2020.} \cite{libraries.io2020} to collect all packages in the npm ecosystem. We filter and label the packages, extract characteristics and derive new features, and use them to train a Random Forest model.

A replication package of our study is available on Zenodo \cite{replication}.

\subsection{Data filtering and labeling}
For this study, we only consider packages with two or more runtime dependents. We want to investigate the most common dependency update strategy for each package. 
Therefore, we should only consider packages that have downstream dependents. Additionally, looking for a majority agreement between dependents of a package is not a sound approach if the package has fewer than 2 dependents. The npm package manager allows developers to specify development dependencies (will be used in development environment) and optional dependencies (npm will try to fetch them but will not raise errors if unsuccessful). We do not consider development and optional dependencies because they are not required for the dependent package to function and are sometimes incomplete.  These thresholds help in removing unused and noisy packages from the dataset. However, we were still able to identify multiple spam packages which had the sole purpose of depending on all packages in npm. The ones we identified were all-packages-X, wowdude-X and neat-X, in all of which the X is replaced by various numbers. 

In order to identify package specialization, we extracted the runtime dependency relationships from the latest published versions of all packages to other packages in our dataset (January 2020). We used the reverse relationship (from the target package to the source package) to determine the dependents of each package and their dependency constraints.
If \rev{more than 50\%} of a package's dependents agree on a dependency update strategy (Section~\ref{sec:Background}), the package is labeled as specialized towards that strategy (i.e. balanced, restrictive, permissive). Otherwise, the package is labeled as unspecialized.  

\begin{figure}[h]
	\centering
	\includegraphics[width=0.85\linewidth]{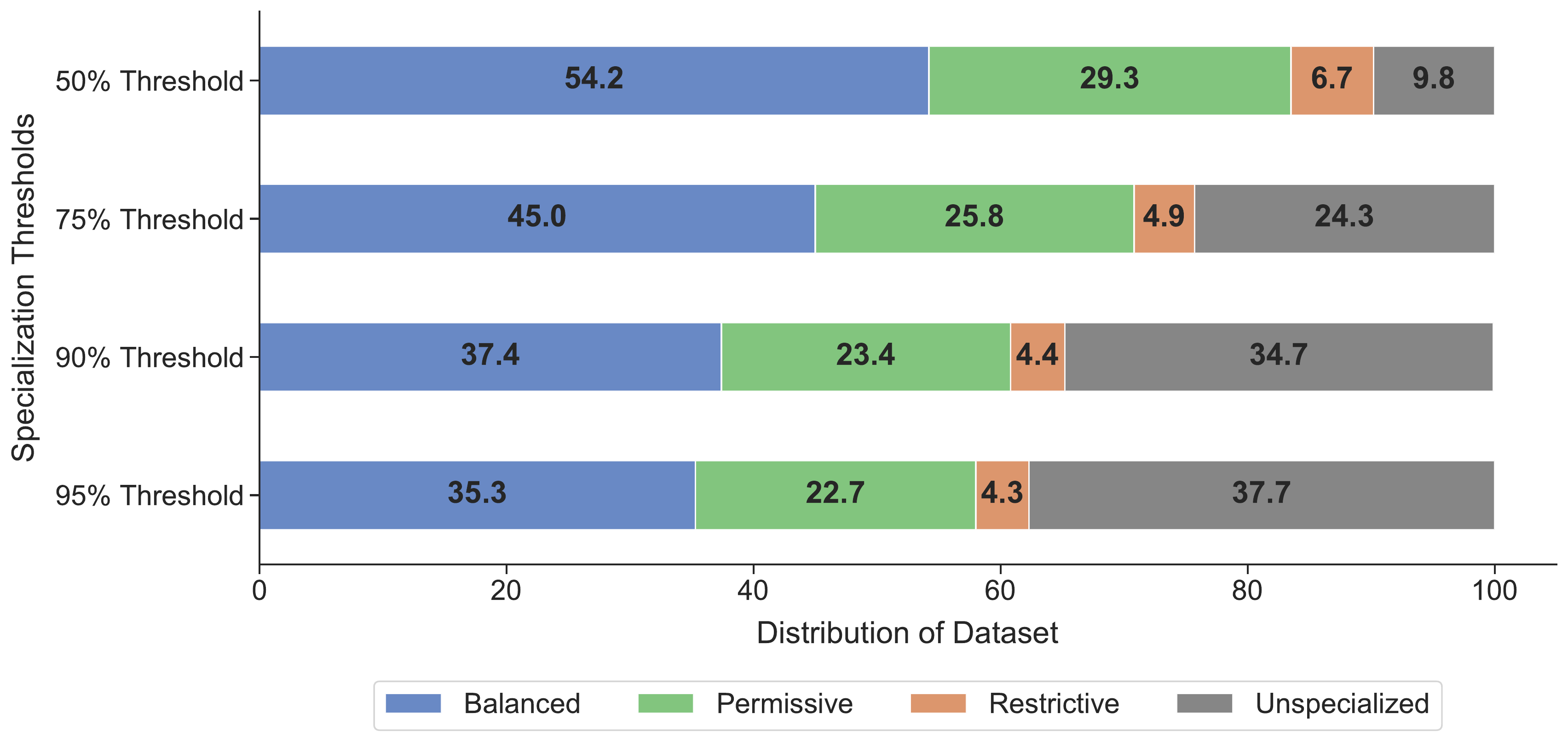}
	\caption{\rev{Impact of specialization threshold on class distribution}}
	\label{fig:stacked-dist}
\end{figure}

This groups all packages in the dataset into 4 categories (balanced, restrictive, permissive, unspecialized). \rev{We do not choose a threshold below 50\% since a threshold of over 50\% for one class is guaranteed to always represent the most accepted update strategy for that package. Increasing the threshold (higher majority agreement) bolsters the confidence in the ``most common update strategy'' when there is an agreement, but as the agreements become rare, the results become less meaningful in practice. As can be seen in Figure~\ref{fig:stacked-dist}, our selected threshold also results in the lowest comparative percentage of ``unspecialized'' packages. Unspecialized packages are not helpful in studying the common update strategy, since by definition, they do not have a common agreed upon update strategy among their dependents.}

\begin{figure}[h]
	\centering
	\includegraphics[width=0.9\linewidth]{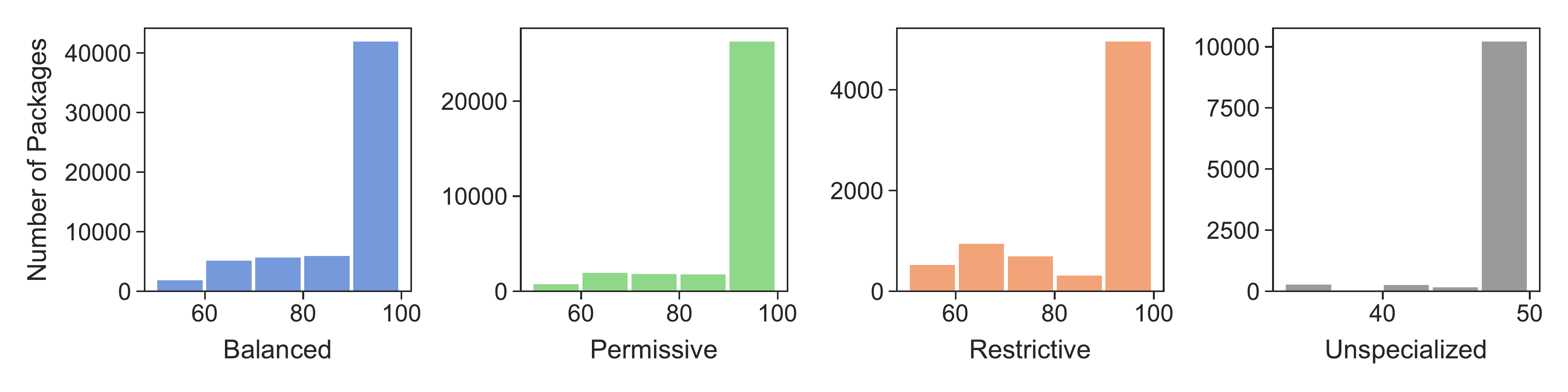}
	\caption{\rev{Distribution of dependent agreement percentage for packages in each class}}
	\label{fig:category-dist}
\end{figure}

The final dataset includes 112,452 total npm packages. From this total, \rev{101,381 (90.2\%)} are specialized toward a particular update strategy and \rev{11,071 (9.8\%)} are unspecialized. Looking at different update strategies we see that \rev{54.2\%} of packages are specialized toward the balanced strategy, 6.7\% are specialized toward the restrictive and \rev{29.3\%} are specialized toward the permissive update strategy. The packages in our dataset have a median of 3 dependents and a median age of 39 months. The distribution of our dataset is shown in the first row (50\% threshold) of Figure~\ref{fig:stacked-dist} \rev{and the distributions of agreement percentage (among dependents) for each class are presented in Figure~\ref{fig:category-dist}.}

\subsection{Feature selection and extraction}

In this section, we explain the rationale for selecting the package features. We then explain our feature extraction procedure and the necessary pre-processing of the features.

\noindent\textbf{Feature selection rationale:}
In order to train a suitable model in predicting dependency update strategies, we first need to select appropriate features that can capture developer needs in choosing the correct strategy. The libraries.io dataset consists of over 50 characteristics for each package, although some are highly correlated. We use the term package features to refer to characteristics from both the package on npm and its project repository. In order to \rev{determine what features in our dataset are relevant and what other features might be needed}, we studied the literature to identify which package characteristics are associated with the characteristics involved in choosing and managing dependencies.

Table~\ref{tab:relevant_features} presents each of these features. All of the studies referenced in the table are comprised of developer surveys and interviews regarding practitioner needs and practices (see Section~\ref{sec:Related Work}). The features listed here are deemed relevant in the literature in choosing and managing dependencies, but ours is the first study to investigate their influence on the dependency update strategy. According to the reviewed literature, developers use the following characteristic groups to select dependencies:

\begin{table}
\small
	\caption{Relevant features in selecting dependencies}
	\label{tab:relevant_features}
	\begin{tabular}{l|r}
		\toprule
		Feature & Studies\\
		\midrule
		Repository Stars Count & \cite{pashchenko2020qualitative,larios2020selecting,haenni2013categorizing}\\
		Repository Watchers Count & \cite{larios2020selecting,haenni2013categorizing}\\
		Repository Forks Count & \cite{larios2020selecting,haenni2013categorizing}\\
		Dependency Count  & \cite{larios2020selecting}\\
		Dependent (Repository and Package) Count & \cite{pashchenko2020qualitative,larios2020selecting,bogart2016break,haenni2013categorizing}\\
		Repository Contributors Count & \cite{pashchenko2020qualitative}\\
		Repository Open Issues Count & \cite{pashchenko2020qualitative}\\
		Licenses & \cite{pashchenko2020qualitative,haenni2013categorizing}\\
		Days Since Last Release & \cite{larios2020selecting,bogart2016break}\\
		Age & \cite{larios2020selecting}\\
		Version Count, Version Frequency & \cite{larios2020selecting,bogart2016break,haenni2013categorizing}\\
		Repository Readme, Description, Wiki, Pages & \cite{larios2020selecting,bogart2016break,haenni2013categorizing}\\
		Repository Size & \cite{larios2020selecting,bogart2016break}\\
		Release Status & \cite{bogart2016break}\\
		\bottomrule
	\end{tabular}
\end{table}

\begin{itemize}
	\item \textbf{Package maturity and popularity} is a recurrent factor in the literature. Prominent projects that are established in the community are a priority in selecting dependencies \cite{bogart2016break,haenni2013categorizing,larios2020selecting,pashchenko2020qualitative}. Characteristics such as Age, Dependent Count, Repository Stars and Forks Count along with Repository size and Contributors count can be used as indicators for established package among the community. We hypothesize that packages with a more established history (whether positive or negative) provide more information for developers to decide on their preferred dependency update strategy. Popular packages are also encouraged to be more diligent in their updates as they are scrutinized by a larger user-base. Additionally, packages in initial stages of development are often deemed unstable by dependency guidelines such as SemVer, and thus warrant stricter update strategies.
	 
	\item \textbf{Package activity and maintenance} is cited as one of the most important factors in selecting dependencies \cite{bogart2016break,larios2020selecting,pashchenko2020qualitative}. Characteristics such as Version frequency, Open issues count and Days since last release can be used as indictors for package activity. We hypothesize that highly active packages would be more problematic for dependents that opt for permissive dependency approaches as the likelihood of breaking changes may increase with more frequent releases. On the other hand, different dependency update strategies can be inconsequential for packages that have not released a new version for a long time as there is little meaningful difference between the latest version and an old version.
	
	\item \textbf{Documentation} is also among the highly stated factors for selecting dependencies \cite{bogart2016break,haenni2013categorizing,larios2020selecting,pashchenko2020qualitative}. License information is also important to prevent legal issues. Project readme and wiki files, along with license information can be used as suitable indicators for this category. \rev{We use the license code as a feature that represents the type of licenses for the package (e.g. MIT, BSD-2-Clause, ISC).} We hypothesize that the resulting perception from better documentation can not only encourage developers to select a package, but also influence the perception of trust on the package. This in turn can sway them to opt for less restrictive update strategies. Adequate documentation may also bring comfort in knowing that the dependent's development team can rectify shortcomings in particular dependency versions.
\end{itemize}

\noindent\textbf{Feature extraction:}
Some of the selected features are directly available in the libraries.io dataset and others are derived using the raw features in the dataset. In the following, we will explain the derived features:
\begin{itemize}
	\item \textbf{Age} is derived using the package's ``created timestamp'' and comparing it against the date the dataset was released (Jan 2020).
	\item \textbf{Version Frequency} is derived by counting the number of releases and dividing it by the package age in months. In cases where the age was zero months, we used version count instead of version frequency. 
	\item \textbf{Dependent Count} for each package is the sum of reverse dependencies (dependents of a package) from the latest version of all packages in the dataset to that package. The dependent count available in libraries.io also includes dependents from old versions of all packages.
	\item \rev{\textbf{Transitive Dependent Count} is the total number of packages in the dependent tree of our package. It is calculated by converting the dependency relationships for each package into a graph and calculating the total ancestors from the selected package.}
	\item \textbf{Dependency Count} is calculated by counting the number of dependencies for the package.
	\item \rev{\textbf{Transitive Dependency Count} is the total number of packages in the dependency tree of our package. It is calculated by converting the dependency relationships for each package into a graph and calculating the total descendants from the selected package.}

	\item \textbf{Release Status} is extracted using the latest version of the package and determines if the package is in initial development (pre-1.0.0) or production stage (post-1.0.0).
	\item \textbf{Days Since Last Release} is derived by extracting the latest release and comparing its date against the date the dataset was released (Jan 2020).
\end{itemize}

We hypothesize that the \textbf{Domain} or type of the package may influence how developers depend on a package since certain dependencies may correspond to more critical aspects of a software project. This is further investigated in the manual analysis of Section~\ref{sec:Results}. Seeing that we have access to package keywords, we can use them to assign domain/type to each package. Since there are many varied keywords in the dataset, we first need to prune the keyword set and map each package to a smaller set of keywords. To this aim, we first address highly correlated keywords by finding the top 2000 trigrams and bigrams (n-grams are collections of n keywords that frequently appear together) with the highest Point-wise Mutual Information (PMI) scores. PMI is a metric provided by NLTK \cite{nltkdoc} to quantify the likelihood of co-occurrence for two words, taking into account that this might be caused by the frequency of single words. We only consider trigrams and bigrams that appear at least 10 times in the dataset. In short, we group keywords into sets if they commonly co-appear. We then use one keyword to represent each set. This procedure reduces the average number of keywords per package. In the next step, we use the keywords to cluster the packages. To this aim, we use the top 15 keywords to build a term frequency vectorizer for package keywords. The vectorized keywords are fed into a K-means clustering algorithm with K=10 (derived using the elbow and silhouette methods \cite{geron2019hands}). The result is a numerical ``Domain'' feature which includes a value from 1 to 10 for each package.

\noindent\textbf{Feature pre-processing:}
Many values in the dataset did not have a default of zero and instead, included missing values. Missing values were handled in such a way that would be meaningful for each feature. For example, if there were missing values for the number of dependencies or repository stars count, a value of zero was used as a replacement. However, this strategy would not be meaningful for all features. For example, missing values in repository size were replaced by the median repository size. Since we study packages with a dependent count greater or equal to 2, missing values in dependent count were automatically removed.

Highly correlated features negatively impact the model's performance and more importantly, its interpretability. We calculate the Pearson correlation and remove features with a correlation above 0.7.

When two features were highly correlated, we kept the feature with the more tangible description. For example "Repository Contributors Count" was removed as it was highly correlated with "Repository Size" and "Repository Watchers Count" was removed due to its high correlation with "Repository Stars Count". \rev{In total, the following 12 features were removed due to correlation: Repository Host Type, Repository Wiki enabled?, Repository Pages enabled?, Repository Open Issues Count, Repository Issues enabled?, Repository Watchers Count, Repository Forks Count, Repository SourceRank, Versions Count, Repository Contributors Count, Repository URL, Transitive Dependent Count.} 

Table~\ref{tab:selected_features} presents the final set of features selected for this study along with a description for each feature. \rev{After dropping the aforementioned correlated features, the remaining feature set in Table~\ref{tab:relevant_features} appears in our final set of features. We have also used the characteristic groups observed in the literature (maturity and popularity, activity and maintenance, and documentation) to utilize relevant features available in the dataset or synthesize relevant features. Transitive dependency count is an extension of dependency count which considers whether the dependencies of a package are "dependency heavy" themselves. The existence of keywords and homepage URL is another means of evaluating package documentation. The domain is an attempt to identify package type by clustering the keywords (since the entire set of keywords are too numerous to use outright). The domain and keywords features have different objectives. Domain attempts to encapsulate package type while the existence of keywords is an indicator of package documentation. License code is also different from repository license in a similar manner. The former is a means of encapsulating package license type and permissions (to understand whether it affects how dependents use the package) while the latter is an indicator of documentation completeness. We also added SourceRank as a feature as it is the scoring algorithm used by Libraries.io to index the results \cite{sourcerank2020}. SourceRank aggregates a number of metrics believed to represent high quality packages, some of which are also included in our features. For example: Is the package new? How many contributors does it have? and Does it follow SemVer?}

\begin{table}
	\small
	\centering
	\caption{Selected features and their description}
	\label{tab:selected_features}
	\begin{tabular}{l|l|c}
		\toprule
		Feature & Description & \rev{Histogram}\\
		\midrule
		Dependency Count & The \# of dependencies from the latest releases of npm packages. & \includegraphics[width=0.1\linewidth]{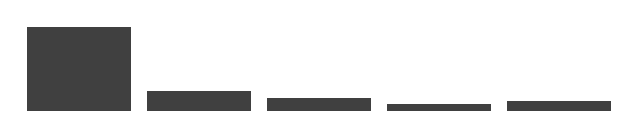}\\ 
		\rev{Transitive Dep. Count} & \rev{The \# of transitive dependencies from the latest package release.} & \includegraphics[width=0.1\linewidth]{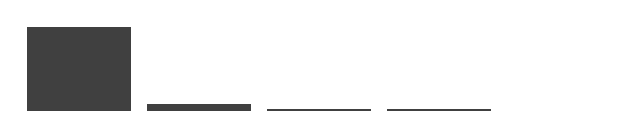}\\ 
		Dependent Count & The \# of dependents from the latest releases of npm packages. & \includegraphics[width=0.1\linewidth]{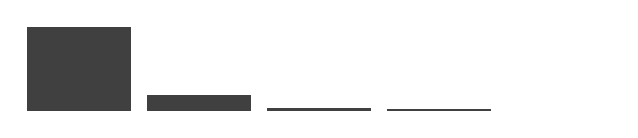} \\
		Version Frequency & The \# of released versions divided by the age. & \includegraphics[width=0.1\linewidth]{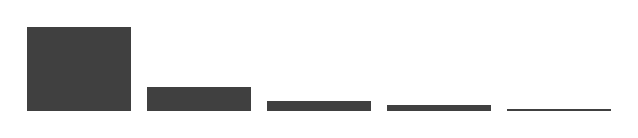}\\
		Age & The age of the project in months. & \includegraphics[width=0.1\linewidth]{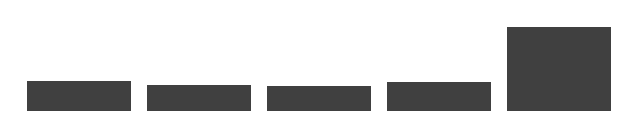}\\
		Description & Whether or not the package provides a description. & \includegraphics[width=0.1\linewidth]{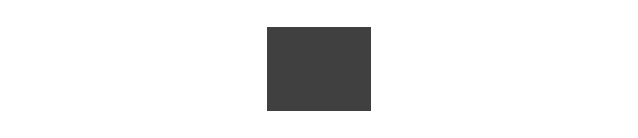}\\
		Keywords & Whether or not the package specifies keywords. & \includegraphics[width=0.1\linewidth]{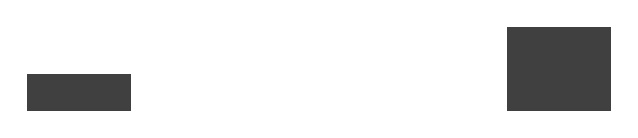}\\
		Homepage URL & Whether or not the package specifies a homepage URL. & \includegraphics[width=0.1\linewidth]{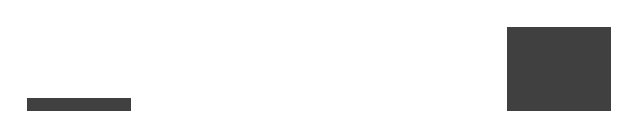}\\
		License Code & The ID for the type of license(s) specified for the package. & \includegraphics[width=0.1\linewidth]{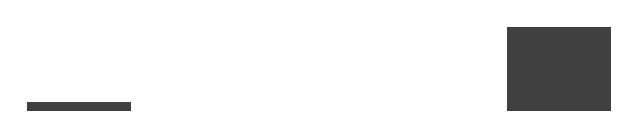} \\
		SourceRank & The SourceRank metric of a package provided by libraries.io. & \includegraphics[width=0.1\linewidth]{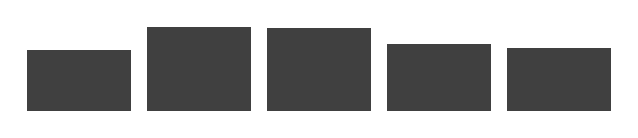}\\
		Release Status & Whether or not the package is at a pre-1.0.0 or post-1.0.0 state. & \includegraphics[width=0.1\linewidth]{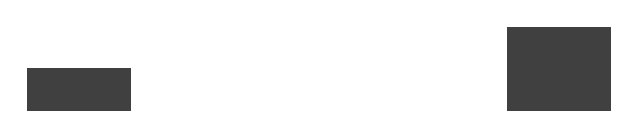}\\
		Days Since Last Release & The \# of months elapsed since the most recent release. & \includegraphics[width=0.1\linewidth]{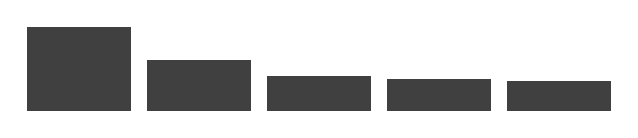}\\
		Dependent Repositories & The \# of dependent repositories on the package's repository. & \includegraphics[width=0.1\linewidth]{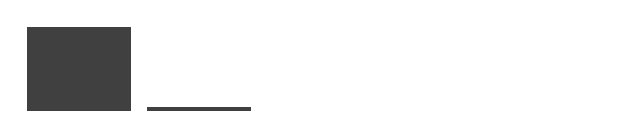}\\
		Repository Size & The size of the package repository in Kilobytes. & \includegraphics[width=0.1\linewidth]{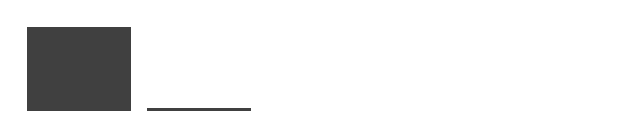} \\
		Repository Open Issues & The \# of open issues in the package repository. & \includegraphics[width=0.1\linewidth]{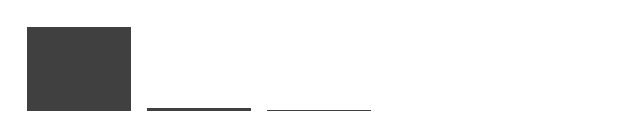} \\
		Repository Stars & The \# of stars for the repository. & \includegraphics[width=0.1\linewidth]{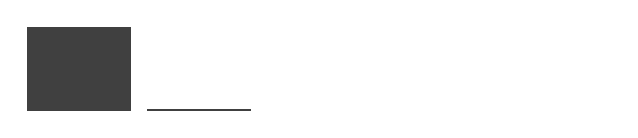} \\
		Repository License & Whether or not the package repository specifies a license. & \includegraphics[width=0.1\linewidth]{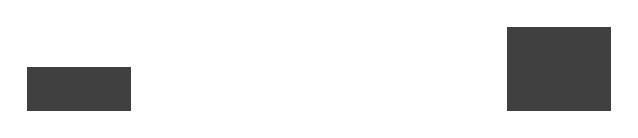}\\
		Repository Readme & Whether or not the package repository provides a readme file. & \includegraphics[width=0.1\linewidth]{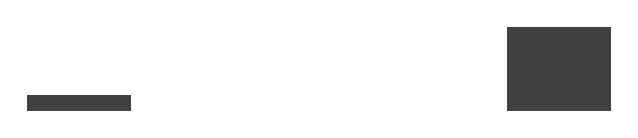}\\
		Domain & Package domain group extracted from the keywords. & \includegraphics[width=0.1\linewidth]{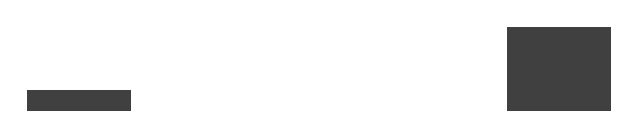}\\

		\bottomrule
	\end{tabular}
\end{table}

\section{Findings of the Study}
\label{sec:Results}
We present the findings of our empirical study starting by our results for using package characteristics to predict the dependency update strategy. This is followed by a study on the impact of package characteristics on the popular dependency update strategy. In the last section of our results, we conduct a mix-method analysis with 160 packages to understand the contributing factors in the evolution of update strategies over a span of 10 years.

\subsection{Can package characteristics be used as indicators of dependency update strategies?}

\noindent
\textbf{Motivation:} Understanding the association between package characteristics and the commonly chosen dependency update strategy by its dependents can help the community to better grasp the dynamics of dependency update strategies. Knowing whether or not the characteristics of a package are indicators of dependency update strategies will also help developers by providing them with meaningful and actionable information in the process of deciding the appropriate update strategy for their \rev{package dependencies}. This can help prevent dependency issues that result from using unsuitable alternative strategies \cite{jafari2020dependency}.

\begin{figure}[h]
	\centering
	\includegraphics[width=0.65\linewidth]{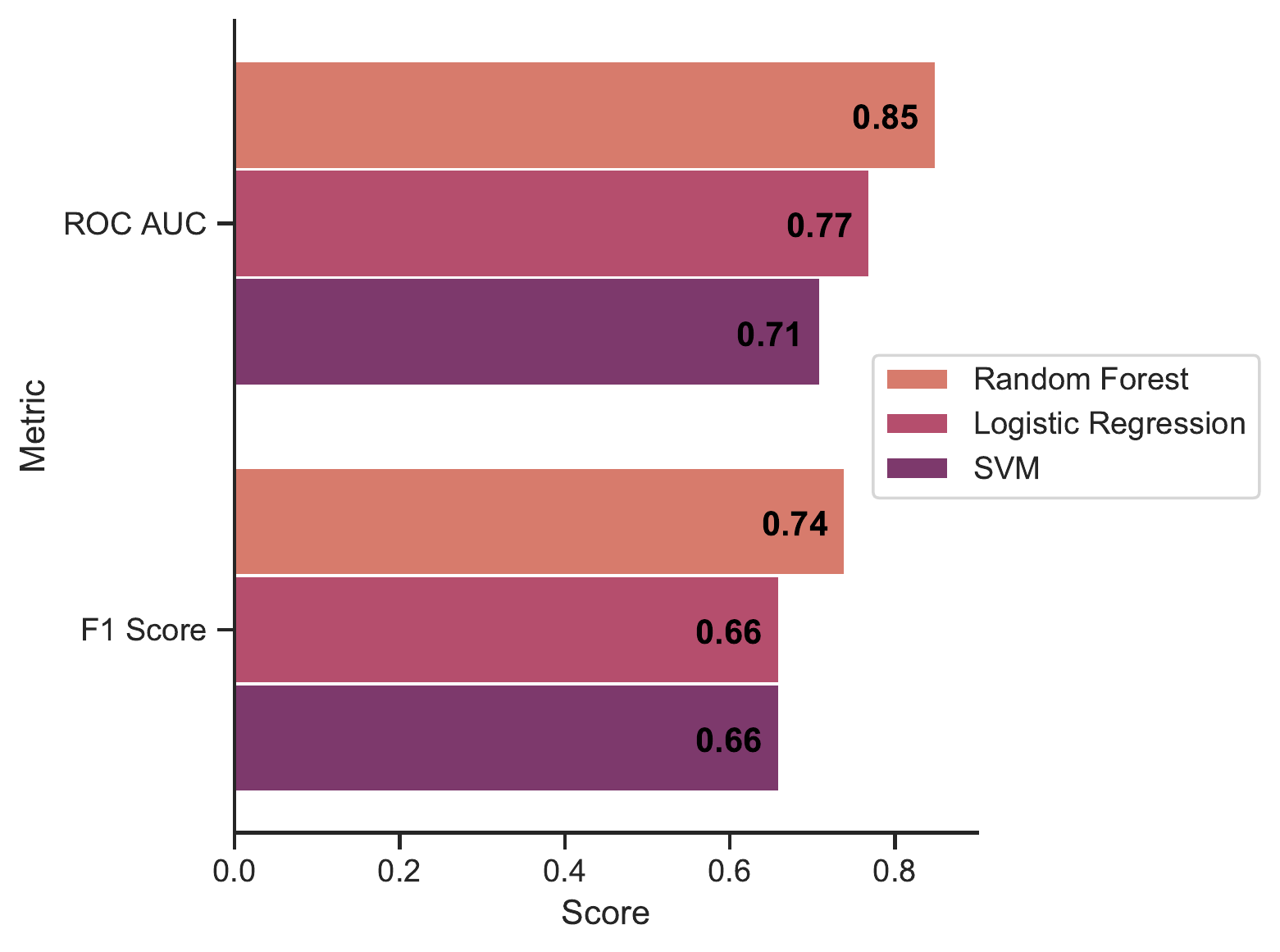}
	\caption{\rev{Comparison of performance for candidate models}}
	\label{fig:models_pre}
\end{figure}

\noindent\textbf{Approach:}
In order to study the relevance of package characteristics to the commonly used dependency update strategy by the community, we use the features in Table~\ref{tab:selected_features} to train a Random Forest model. The \rev{multi-class} model aims to use the characteristics to predict the commonly used update strategy for each package. \rev{The result of the prediction for each package can be one the four classes of Balanced, Restrictive, Permissive or Unspecialized. The unspecialized class does not represent an update strategy but rather, packages which do not have a common agreed-upon update strategy among their community of users.} We use Random Forests since the objective of our study is to understand the association between package characteristics and dependency update strategies which necessitates descriptive models. In addition, we want good performance compared to the baseline in order to derive meaningful associations. \rev{We conducted preliminary experiments with Random Forest, Logistic Regression and SVM and compared their performance using ROC-AUC and F1-score metrics. The ROC (Receiver Operating Characteristics) is a probability curve where AUC (Area Under the Curve) is a value between 0 and 1 that represents the degree of which the model is capable of distinguishing between classes. The higher the AUC, the better the model is at correctly predicting classes. Since our problem is a multi-class model, we plot multiple ROC-AUC curves, one for each of the classes using the One-vs-Rest (OvR) methodology. The final ROC-AUC is the resulting average of the ROC-AUC scores. F1-score is a function between 0 and 1 that balances between precision (the fraction of true positive instances among the retrieved instances) and recall (the fraction of true positive instances that were retrieved). 
We did not modify the hyper-parameters of the three models but we performed data normalization which is important for Logistic Regression and SVM when there is high cardinal variance between the features. All three models were trained on 80\% of our dataset (training set) and evaluated on the held-out 20\% (tests set). As can be seen in Figure~\ref{fig:models_pre}, the Random Forest model yields considerably better performance, which is why it is selected as the Package Characteristics model in this study.} 

Since there is no previous work on using package characteristics to predict dependency update strategies, the results are compared against two baselines; the stratified baseline model and the balanced model. The stratified baseline uses the class distribution in the training set for weighted random predictions about the suitable update strategy. The balanced baseline always predicts the balanced update strategy, as is recommended by npm \cite{npmsemver}. 
We evaluate the performance of the model using ROC-AUC and F1-score metrics \rev{(as explained previously in our preliminary experiments)}. We use 80\% of the data as our training set and leave the remaining 20\% for the final evaluation. We tuned the hyper-parameters of the Random Forest model using 10-fold validation on the training set which results in 500 estimators (trees) with 8 minimum samples required for a split. The 10-fold cross validation fits the model 10 times, with each fit being performed on a 90\% of the training data selected at random, with the remaining 10\% used as a validation set. It is important to evaluate the model on the 20\% of the data used as a held-out set since we want to assess the model's performance on unseen data. 

\begin{figure}[h]
	\centering
	\includegraphics[width=0.85\linewidth]{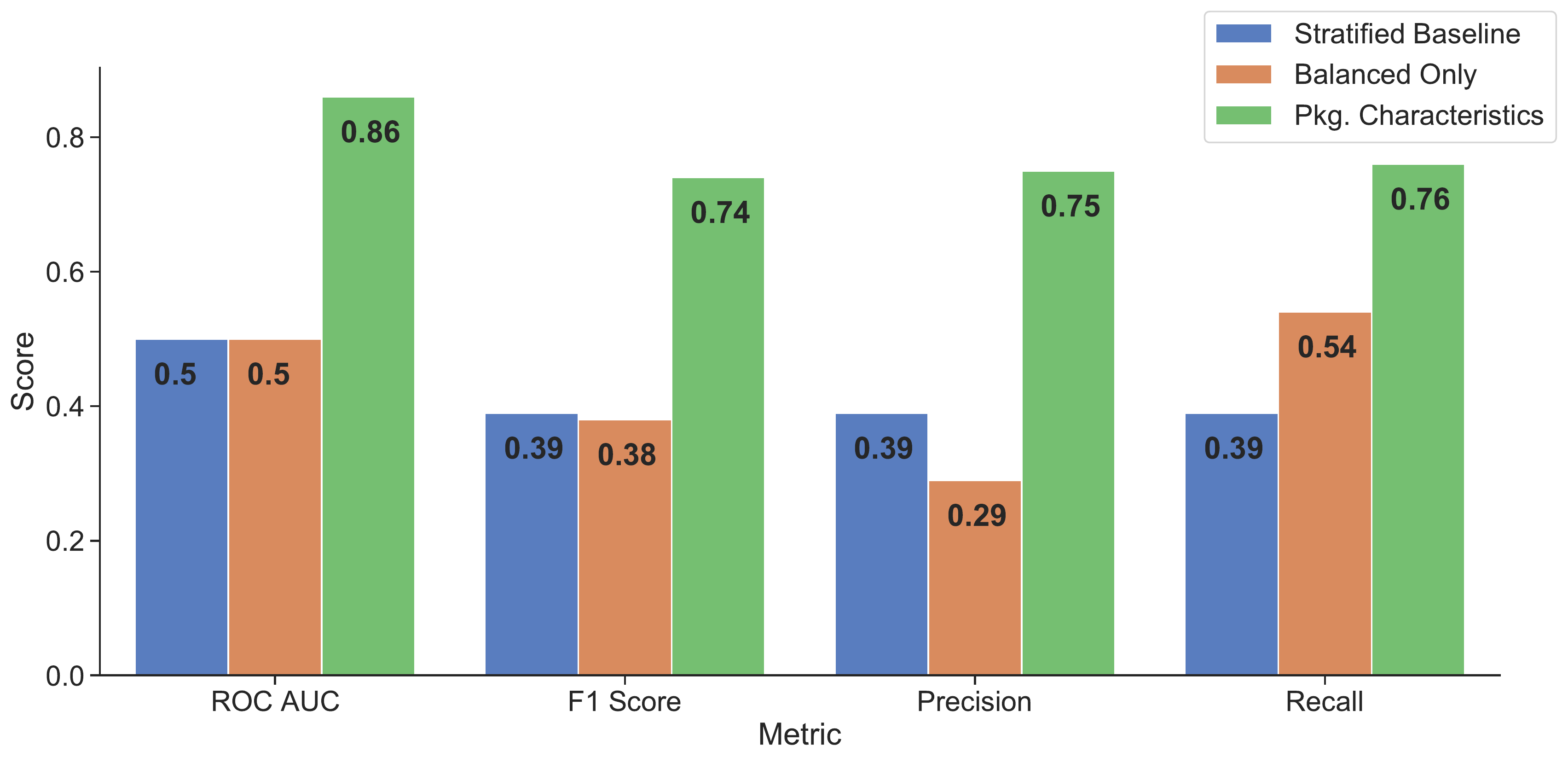}
	\caption{\rev{Performance evaluation results}}
	\label{fig:eval_results}
\end{figure}

\noindent
\textbf{Results:}
Figure~\ref{fig:eval_results} \rev{presents the evaluation results using the ROC-AUC, F1-score, Precision and Recall metrics.} Compared to the baseline model, we can see a 72\% improvement in the ROC-AUC for the Random Forest model, achieving an ROC-AUC of 0.86. The ROC-AUC for the Stratified baseline and the balanced-only approach round-up to 0.5, which is the expected behavior of ROC-AUC when the model makes random predictions or always predicts the same class. We also see a 90\% improvement in the F1-score for the Random Forest model compared to the stratified baseline model, achieving a score of 0.74. Since the real world contains unspecialized cases where no agreement is observed, we have also included these unspecialized packages in the training and evaluation of our model. 

The high ROC-AUC score of 0.86 shows that the package characteristics in Table~\ref{tab:selected_features} are not only relevant for selecting dependencies, but they can also be leveraged to predict the dependency update strategy opted by the majority of developers. In other words, they can be used as indicators of dependency update strategies. Another interesting observation are the results for the balanced baseline. While the balanced strategy is the recommended default by the npm ecosystem \cite{npmsemver}, the results indicate that there is a considerable number of packages for which developers do not believe the balanced update strategy to be suitable.

\rev{In Section~\ref{sec:Methodology and Data}, we discussed the impact of alternative specialization thresholds on the class distribution. Additionally, we have analyzed the impact of alternative specialization thresholds on the performance of our model in Table~\ref{tab:alternative-thresholds}. We look at the change in the ROC AUC and F1-score metrics and also calculate the minimum increase in model performance (i.e. the performance compared to the highest value among the stratified and the balanced only models). As can be seen in Table~\ref{tab:alternative-thresholds}, increasing the specialization threshold to focus on higher majority agreements (i.e. 75\%, 90\%, 95\%) actually results in a more performant model (when comparing each model to the corresponding baselines). However, as stated in Section~\ref{sec:Methodology and Data}, higher specialization thresholds result in a higher number of unspecialized packages for which there is no majority agreement on the update strategy. Our objective is to model the relationship between package characteristics and the common update strategy of its dependents in the npm ecosystem. A model that assumes a strictly high level of agreement among the dependents will be of limited use in practice as such agreement does not exist for many npm packages.}

\begin{table}
	\small
	\caption{Comparing model performance across different specialization thresholds}
	\label{tab:alternative-thresholds}
	\begin{tabular}{cl|cc|cc}
		\toprule
		\textbf{Threshold}	& \textbf{Model} &  \textbf{ROC AUC} & \textbf{Min. Increase} & \textbf{F-1 Score} & \textbf{Min. Increase}\\
		
		\midrule
		\multirow{3}{*}{\textbf{50\%}}		
		& Stratified Baseline 		&	0.50 & - 	& 0.39 	& -	\\
		& Balanced Only 				& 	0.50 & - 	& 0.38 	& - \\
		& Package Characteristics 	&	0.86 & 72\% & 0.74 	& 90\%\\

		\midrule
		\multirow{3}{*}{\textbf{75\%}}		
		& Stratified Baseline 		&	0.50 & - 	& 0.33 	& -	\\
		& Balanced Only 				& 	0.50 & - 	& 0.28 	& - \\
		& Package Characteristics 	&	0.85 & 70\% & 0.67 	& 103\%\\

		\midrule
		\multirow{3}{*}{\textbf{90\%}}		
		& Stratified Baseline 		&	0.50 & - 	& 0.32 	& -	\\
		& Balanced Only 				& 	0.50 & - 	& 0.20 	& - \\
		& Package Characteristics 	&	0.86 & 72\% & 0.68 	& 113\%\\

		\midrule
		\multirow{3}{*}{\textbf{95\%}}		
		& Stratified Baseline 		&	0.50 & - 	& 0.32 	& -	\\
		& Balanced Only 				& 	0.50 & - 	& 0.18 	& - \\
		& Package Characteristics 	&	0.88 & 76\% & 0.70 	& 119\%\\
		\bottomrule
	\end{tabular}
	\color{black}
\end{table}

\conclusion{Finding \#1: \rev{The quality of our classification model shows that package characteristics can be used as indicators of the common update strategy chosen by the package's dependent community.}}

\conclusion{Finding \#2: While the balanced update strategy is recommended by npm, \rev{the recommended update strategy from the package characteristics model is better aligned with the update strategy selected by npm developers.}}

\subsection{Which package characteristics are the most important \rev{indicators for dependency update strategies?}}

\noindent
\textbf{Motivation:} 
There is a large array of characteristics for packages in the npm ecosystem and some create extraneous noise in understanding and selecting the appropriate update strategy while others might even mislead the community. By identifying and studying the most important characteristics that are associated with update strategies, the community can better understand the type of packages that fall into each of the three specialization groups. As previously stated, opting for the suitable dependency update strategy for a package can prevent dependency issues that arise from alternative update strategies \cite{jafari2020dependency}. Therefore, developers also need to know which characteristics should be prioritized when deciding on an update strategy and how the increase or decrease of such characteristics would impact the commonly selected dependency update strategy.

\begin{figure}[tbh!]
	\centering
	\includegraphics[width=0.8\linewidth]{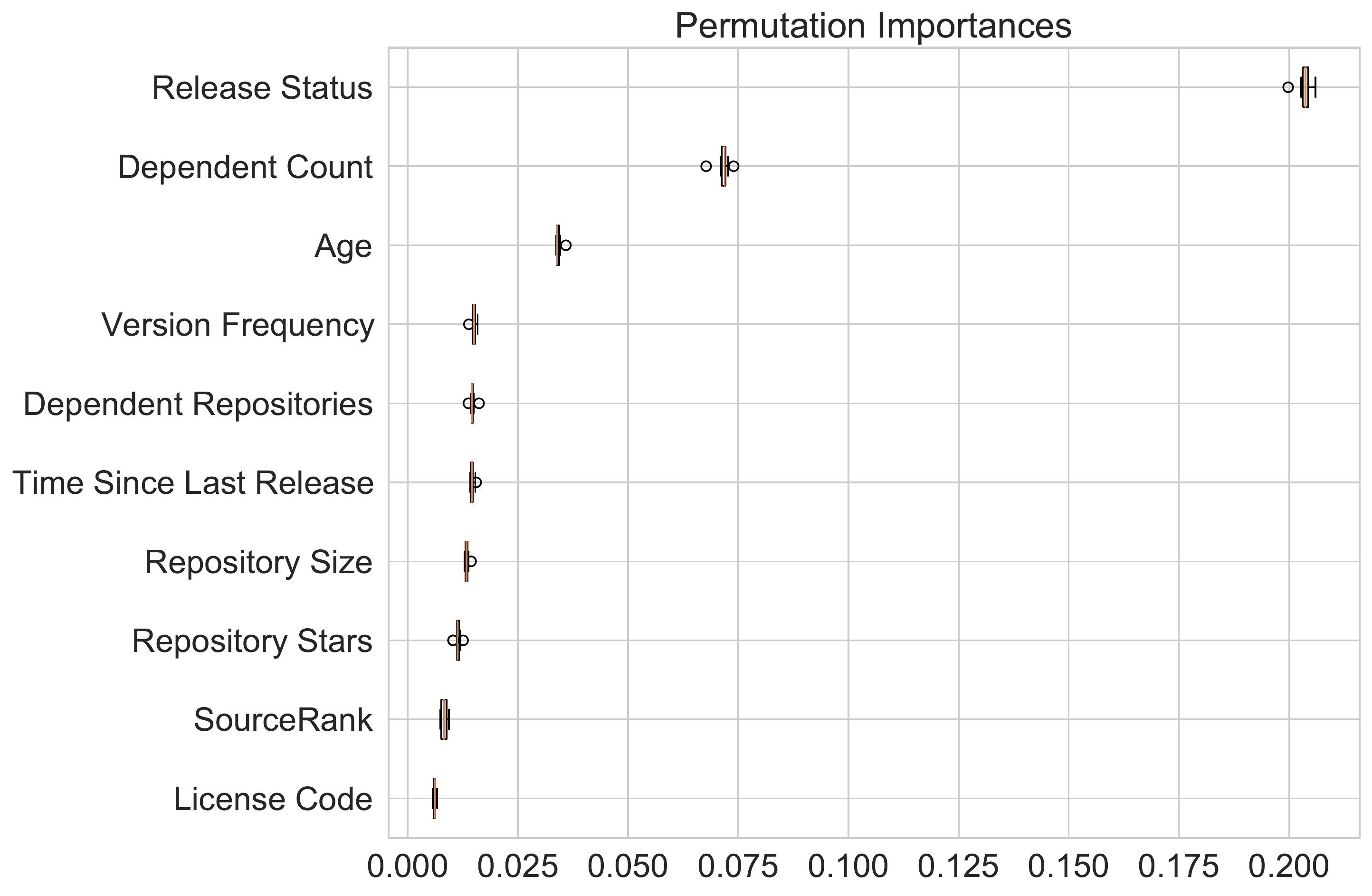}
	\caption{Importance of Features}
	\label{fig:permutation_importance}
\end{figure}

\noindent\textbf{Approach:} 
Package characteristics which have a larger impact on the model's prediction of the commonly used dependency update strategy are better indicators of the update strategy. In order to calculate the feature importance in our model, we use the permutation feature importance instead of the default impurity-based feature importance of Random Forest. The \rev{impurity-based} feature importance inflates the importance of high cardinality features and it is biased to the importance of features in training the model, rather than their capacity to make good predictions \cite{featureimportance}. The 10-fold permutation importances in Figure~\ref{fig:permutation_importance} are calculated by randomly permuting each feature 10 times and observing its impact on the model's performance (ROC-AUC score). A feature is deemed more important if permuting its values has a larger impact on the model's performance.

\begin{figure}[tbh!]
	\centering
	\includegraphics[width=0.8\linewidth]{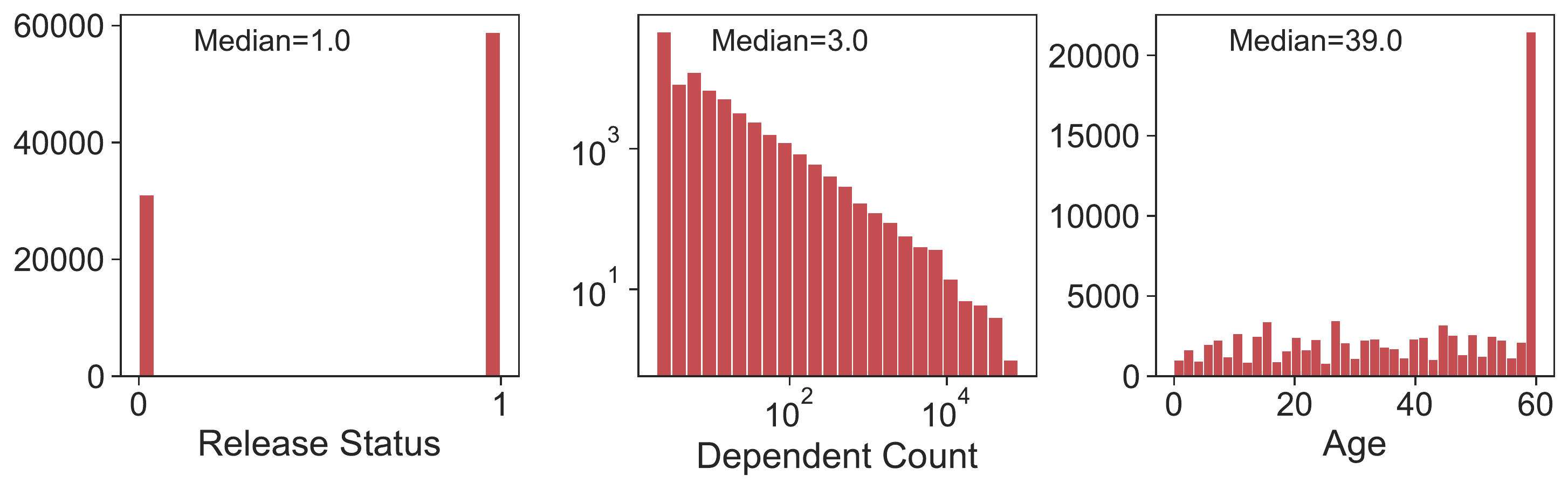}
	\\
	\caption{\rev{Distribution of the top 3 important features (Dependent Count is log10 distribution)}}
	\label{fig:dist_features}
\end{figure}

\begin{figure}[tbh!]
	\begin{subfigure}[tbh!]{\linewidth}
		\centering
		\includegraphics[width=0.8\linewidth]{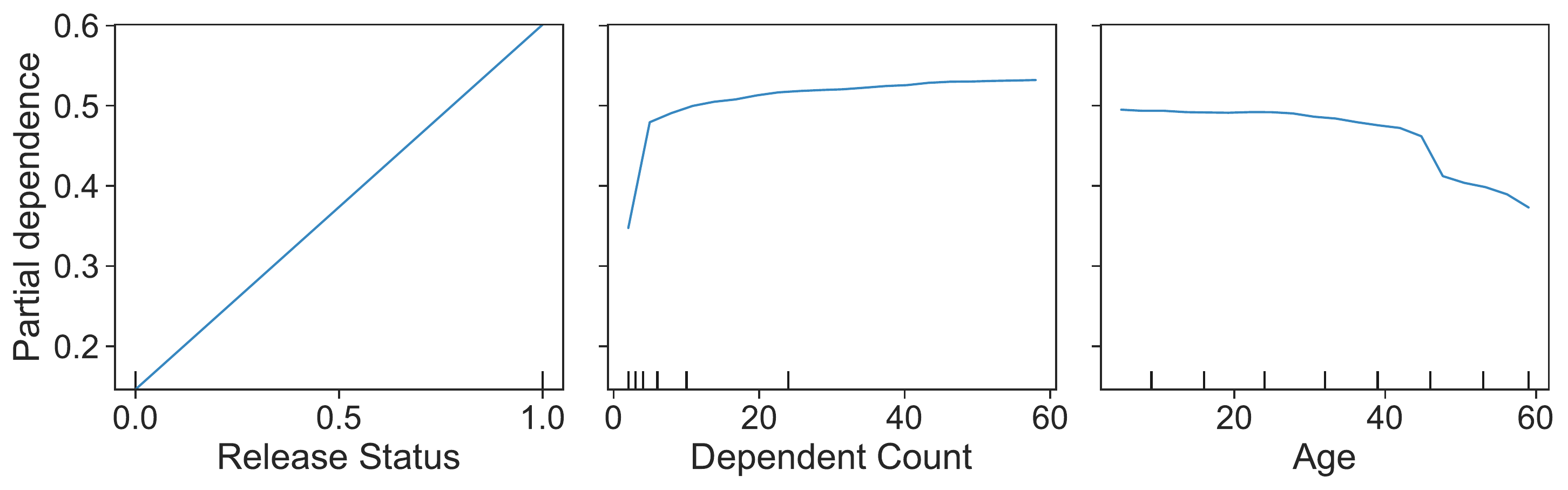}
		\caption{Balanced Class}
	\end{subfigure}
	\begin{subfigure}[tbh!]{\linewidth}
		\centering
		\includegraphics[width=0.8\linewidth]{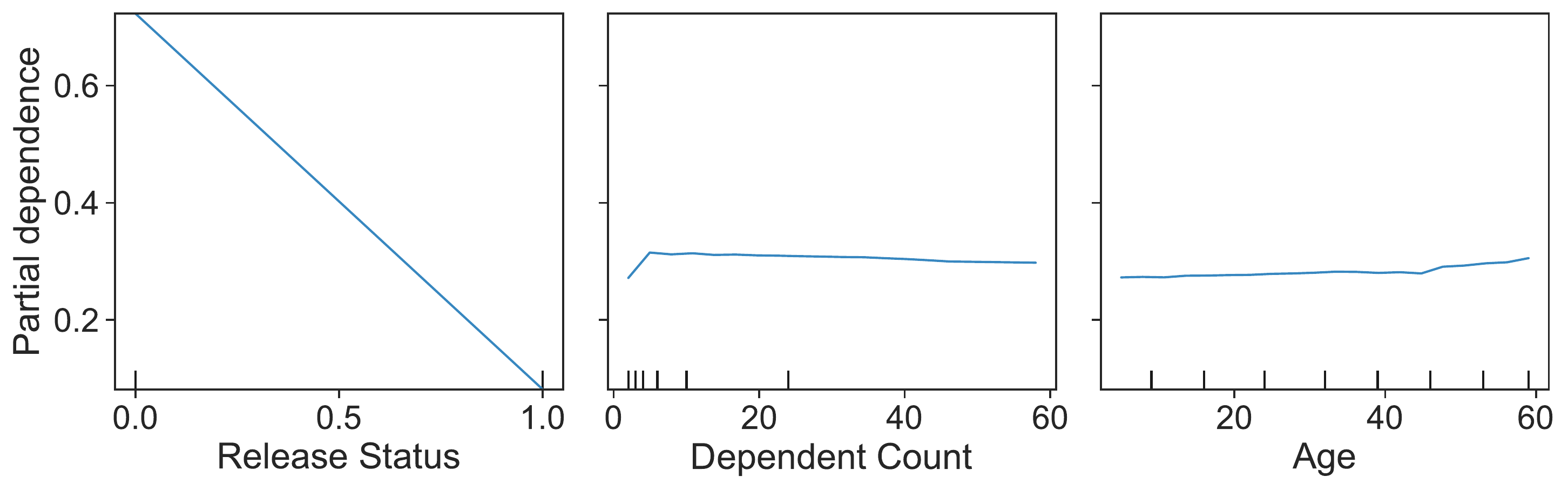}
		\caption{Permissive Class}
	\end{subfigure}
	\begin{subfigure}[tbh!]{\linewidth}
		\centering
		\includegraphics[width=0.8\linewidth]{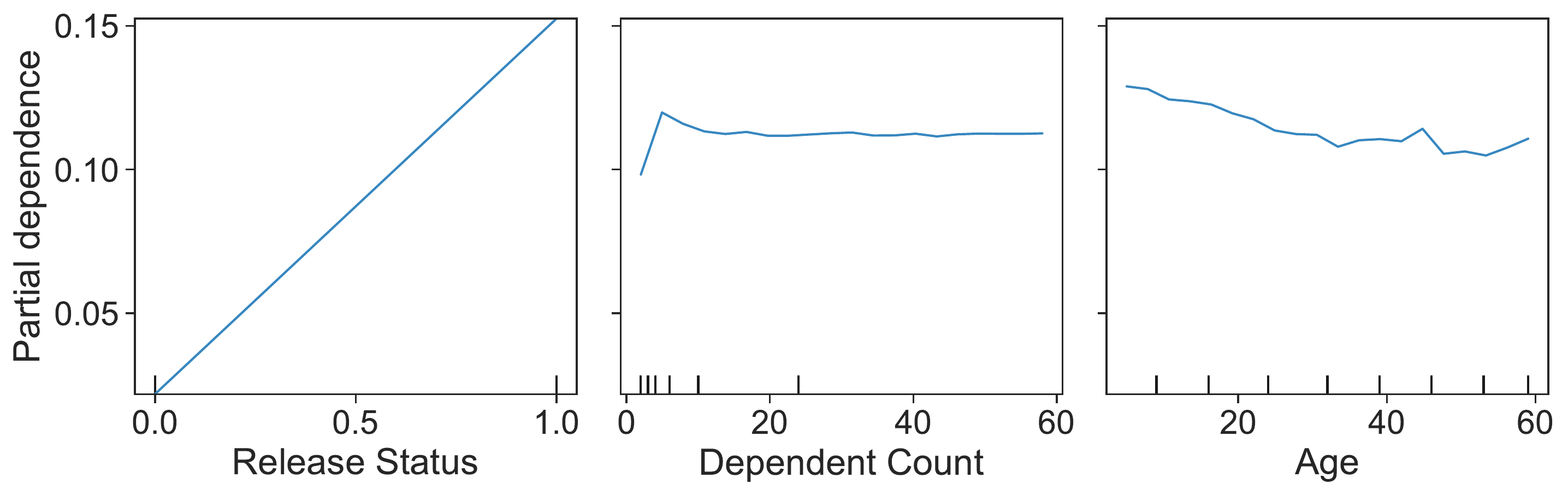}
		\caption{Restrictive Class}
	\end{subfigure}
	\begin{subfigure}[tbh!]{\linewidth}
		\centering
		\includegraphics[width=0.8\linewidth]{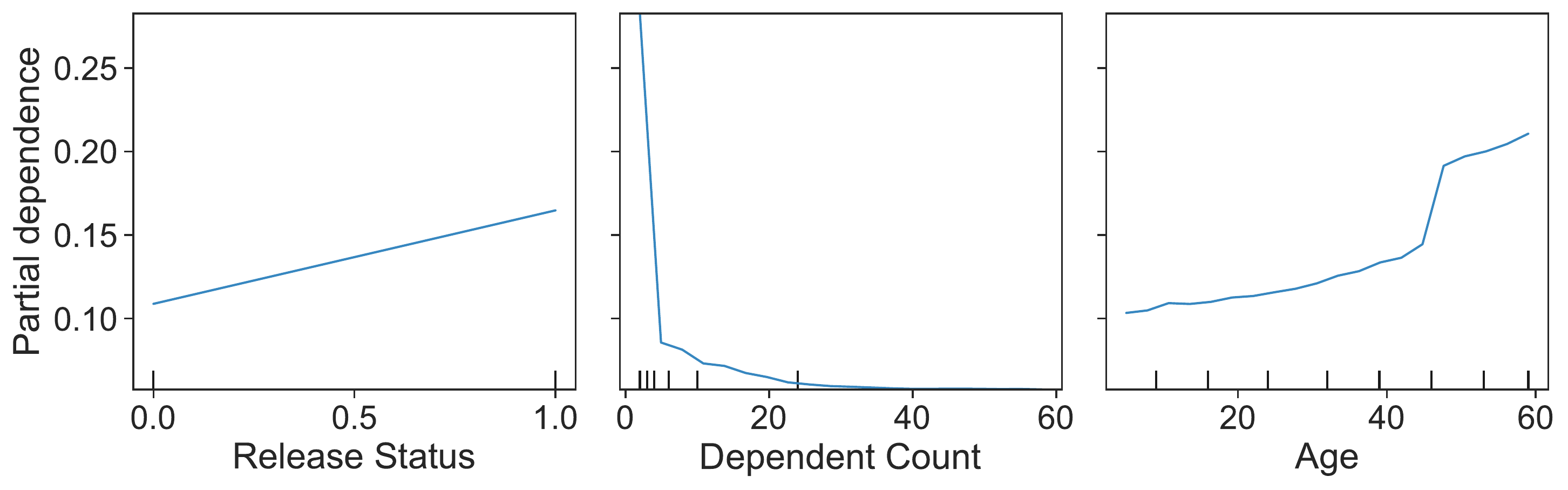}
		\caption{Unspecialized Class}
	\end{subfigure}
	\caption{\rev{Partial Dependence Plots (PDP) for each class}}
	\label{fig:pdp_ice}
\end{figure}

In order to visualize how a change in a package characteristic (feature) impacts the model's decision making for each class, we present Partial Dependence Plots (PDP) for the top 3 important features \rev{in Figure~\ref{fig:pdp_ice}} (since the top 3 are the most prominent). Partial dependence plots visualize the marginal effect of a feature on the prediction of the machine learning model \cite{molnar2020interpretable}. PDPs can highlight linear, monotone or more complex relationships between the feature and the target. In the case of our model, the PDPs \rev{in Figure~\ref{fig:pdp_ice}} can show how an increase or decrease in a feature (such as age) can increase or decrease the model's likelihood to predict the balanced class (or any other class). Since partial dependence is plotted across the distribution, we also plot the distribution plots of the top 3 features to emphasize where the PDPs have more weight. \rev{The Y-axis represents the predicted probability for an instance belonging to the mentioned class.} The tick marks on the X-axis of the PDPs represent the deciles of the feature values, which are consistent with the distributions in Figure~\ref{fig:dist_features}. 

\noindent\textbf{Results:}
The box-plots of Figure~\ref{fig:permutation_importance} present the top 10 most important features which are associated with the commonly used dependency update strategy. As can be seen, release status, dependent count and package age are the most important indicators for dependency update strategies. This hints that these features are highly relevant in influencing decisions about dependency update strategies. Release status is the most relevant feature for the model. Knowing if a package is in early development or post-production is one way to gauge the stability of new releases, which in turn is a way to gauge the degree of freedom dependents give to automatic updates for that package. Additionally, since SemVer considers pre-1.0.0 versions to be unstable, any update strategy that permits even the smallest degree of freedom in receiving new versions (i.e. only accepting patch releases) is considered permissive. This allows the model to use release status to identify many instances of permissive-labeled packages. The high rankings of dependent count and age hints that both popularity and maturity are good indicators of the common dependency update strategy toward the package. 

\conclusion{Finding \#1: The most important indicators for the common dependency update strategy toward a package are its release status, number of dependents and age.}

The distributions for the top 3 features can be seen in Figure~\ref{fig:dist_features}. The majority of packages \rev{(65.5\%)} are in a post-1.0.0 release state with a median of 3 dependent packages and 39 months (3+ years) of age. The distribution of values for most of the top features are highly skewed. Therefore, it is necessary to consider this skewed distribution when analyzing the impact of features.

Figure~\ref{fig:pdp_ice} depicts the partial dependence plots for the top 5 features. The partial dependence plot for release status is unsurprisingly linear since release status is a binary feature. The steep slope of the release status dependence plot is also expected as we previously discovered this feature to be highly important for the model. The impact of release status on the common dependency update strategy is straightforward and intuitive. \textbf{Post-1.0.0 releases result in balanced dependency update strategies, and pre-1.0.0 releases result in more permissive update strategies}. In other words, knowing whether a package is in post-1.0.0 production or in pre-1.0.0 initial development is a good way to decide how permissive or restrictive one should be when depending on that package. As stated previously, this is partly due the treatment of pre-1.0.0 release by the SemVer standard. SemVer considers pre-1.0.0 versions to be unstable by nature and any update strategy that permits even the smallest degree of freedom in receiving new versions (i.e. only accepting patch releases) could introduce backward compatibility issues \cite{semver2019}. \rev{This finding also aligns with the previous investigations of Decan et al. that found the majority of dependencies toward pre-1.0.0 releases to accept patch releases, which is more permissive than what SemVer recommends \cite{decan2021lost}.}

Looking at the partial dependence plots for dependent count, we see that \textbf{higher dependent count increases the likelihood of balanced update strategies } (i.e. dependents of a package tend to agree on the balanced strategy, when the package has more dependents). In a developer survey, Bogart et al. found that the value of avoiding breaking changes grows with the user base of a package \cite{bogart2016break}. Consequently, the user base of such packages may be more likely to perceive the balanced update strategy to be ``good enough'' in preventing breaking changes for highly used and mature packages.

The distribution in Figure~\ref{fig:dist_features} should be taken into account when discussing the PDPs. Since the median dependent count is 3, the left portion of the plot has more weight. It is also important to highlight that \textbf{packages with very few dependents (less than 5) have a considerably higher chance of not being specialized} (i.e. dependents of packages with few dependents are less likely to agree on a dependency update strategy). This is a natural consequence of lesser dependents as there is not yet enough dependents (and perhaps package history) to reach an agreement on how to treat that package as a dependency. Additionally, dependents may be more inclined to choose an update strategy based on personal preference if there is no established popular update strategy for the upstream package.

The partial dependence plots for age reveals that developers tend not to favor the balanced update strategy for old packages, specifically those older than 45 months. Cross referencing this information with the distribution gives further insight. Since the majority of the packages in the dataset are in fact more than 39 months old (right portion of plot has more weight), \textbf{we can conclude that in general, dependents of newer packages favor the balanced update strategies more than dependents of older packages.} The SemVer caret notation was established as the npm default in 2014 \cite{decan2019package,nodesource}. This alone could gradually shape the update strategy the majority of developers choose for newer packages. On the other hand, some might deem an old project as stagnant and will not worry about a new release that breaks the API, which can justify permissive update strategies.

\conclusion{Finding \#2: Package characteristics are highly skewed and packages with less than 5 dependents are less likely to be specialized toward a particular dependency update strategy.}

\conclusion{Finding \#3: Dependents of younger, post-1.0.0 release packages with more dependents are more likely to use the balanced update strategy while dependents of pre-1.0.0 release packages are more likely to use the permissive  update strategy.}

\subsection{How do dependency update strategies evolve with package characteristics?}

\noindent\textbf{Motivation:}
According to our model, characteristics such as release status, dependent count and age have the largest impact on the dependency update strategy. Interestingly, all of these top characteristics are indicative of how a package evolves over time (since dependent count generally increases over time and release status is changed once in a package's lifetime). Consequently, there can be multiple explanations for how the evolution of a package impacts the update strategy chosen by its dependents. For example:

\begin{itemize}
	\item The common update strategy was different early on but dependents gradually shifted to a new update strategy.
	\item The common update strategy changed because new dependents are adopting a different strategy than old dependents.
	\item The common update was initially the same and dependents (new and old) simply followed the previous choice.
	\item The common update strategy experienced a shift due to the shift from a pre-1.0.0 version to a post-1.0.0 version.
	\item The common update strategy experienced a sudden shift due to an anomalous event in the package's lifecycle.
\end{itemize}
 
 While we know that release status, dependent count and age are related to the currently popular dependency update strategy, we need to see if such a relationship was preserved through the package's evolution or if perhaps, it is a result of an external event. Understanding the evolution of dependency update strategies toward a package will provide much needed insight into why the characteristics that are most relevant to the dependents' update strategy are all related to a package's evolutionary behavior. 

\noindent\textbf{Approach:}
Evaluating the evolution of dependency update strategies is carried out through a mix of quantitative and qualitative techniques. We take a random sample of 160 packages from the dataset (40 packages from each of the three update strategies + 40 unspecialized packages) for a historical analysis of each package's dependents over the last 10 years up to the latest snapshot of the dataset. We want to look at packages with over 100 dependents in the hopes of disregarding packages with very limited historical dependent data. Therefore, half of this sample dataset consist of packages with 100 to 1000 dependents (in the latest snapshot) and the other half have more than 1000 dependents (in the latest snapshot). This sample of 160 packages is not meant to be a representative sample of the main dataset. Rather, it is ``convenience sample'' \cite{baltes2022sampling} consisting of reasonably used packages selected for an in-depth mix-method study that is otherwise not feasible on a large dataset.

For each package, we utilize a monthly snapshot of the ecosystem to identify dependents at each month. We then analyze the dependency requirement constraints to identify the number of dependents using a particular update strategy per month. Since the age of a package increases with time, visualizing the dependency update strategies over time is akin to plotting the evolution of update strategies over the package's lifecycle. It is important to note that even though we take 40 samples from each group (balanced, restrictive, permissive, unspecialized), we still plot all update strategies for each package, since a package currently specialized toward a restrictive update strategy for example, may have other strategies used by its dependents throughout time. To eliminate the bias toward dependents that release more frequently, we only consider the latest version of each dependent at each month (i.e. each dependent package is counted only once per month, regardless of how many versions it maintains).

\noindent\textbf{Results:}
We present the commonly observed evolution patterns for dependency update strategies along with real examples that embody the findings. While age and dependent count do not increase at the same rate, their relationship with the evolution of update strategies proved to be similar. Thus, we focus our analyses on the evolution of dependency update strategies across package age. The complete set of visualizations for each package can be accessed through our replication package \cite{replication}.

\begin{figure}[tbh!]
	\begin{subfigure}[b]{0.5\linewidth}
		\includegraphics[width=\linewidth]{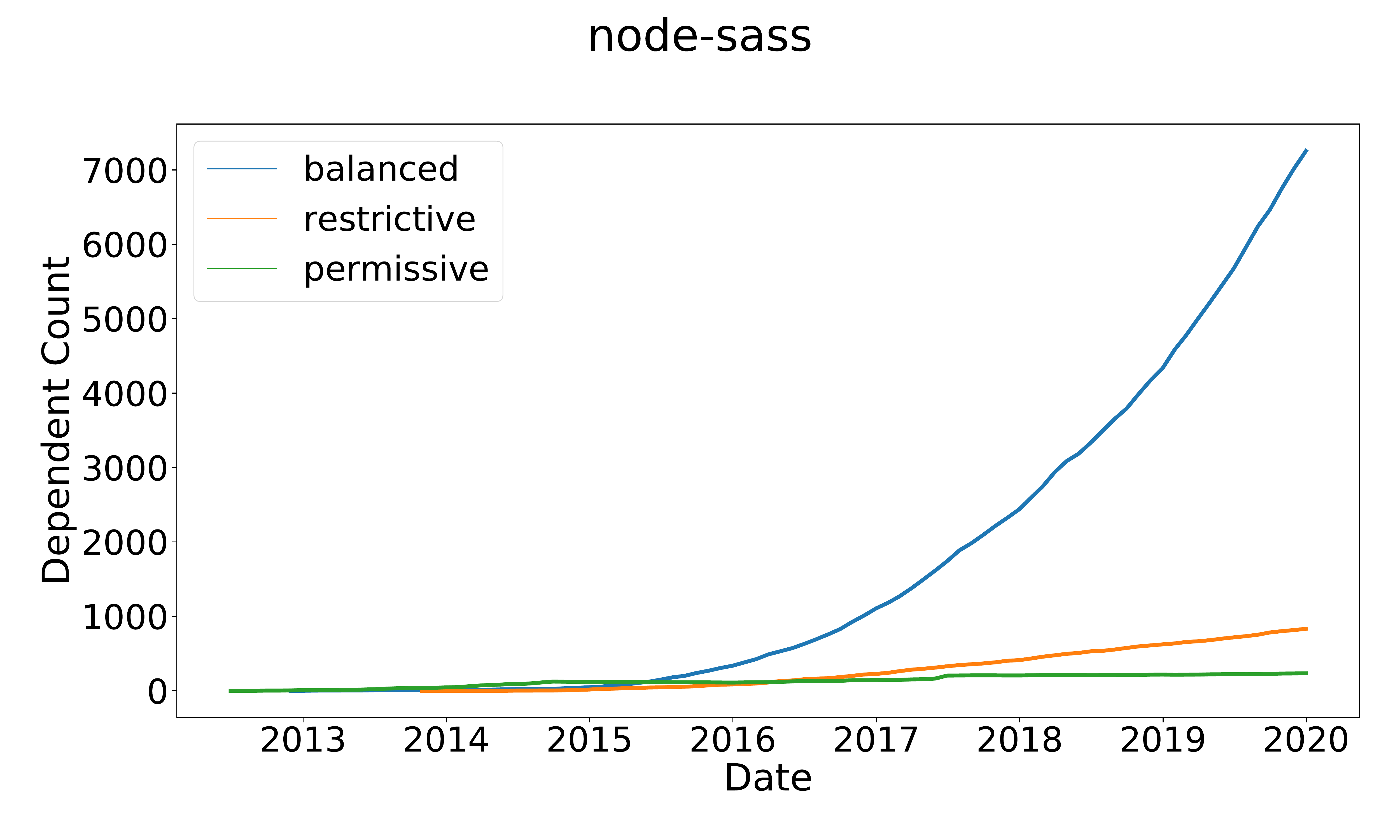}
	\end{subfigure}
	\hspace{-1mm}%
	\begin{subfigure}[b]{0.5\linewidth}
		\includegraphics[width=\linewidth]{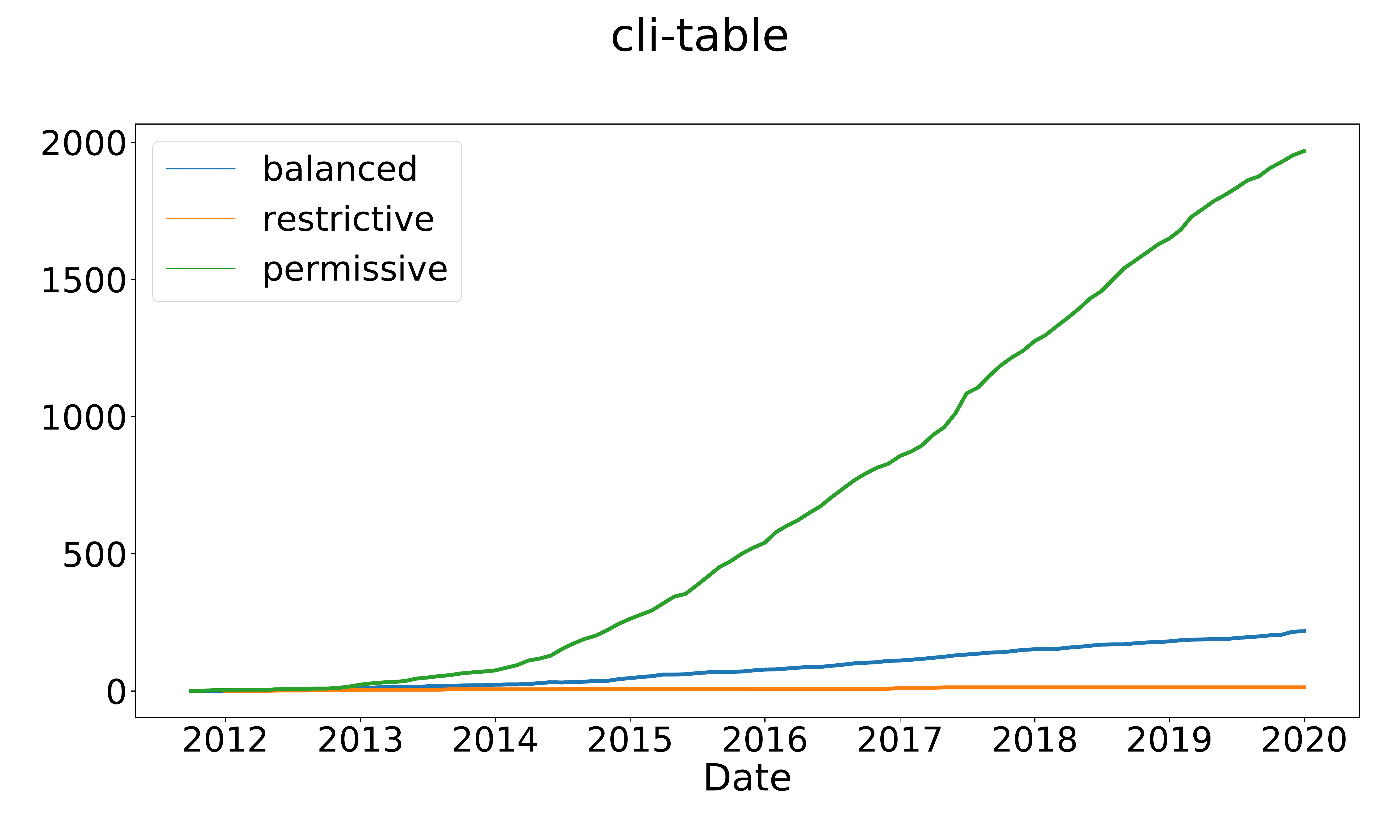}
	\end{subfigure}
   \begin{subfigure}[b]{0.5\linewidth}
		\includegraphics[width=\linewidth]{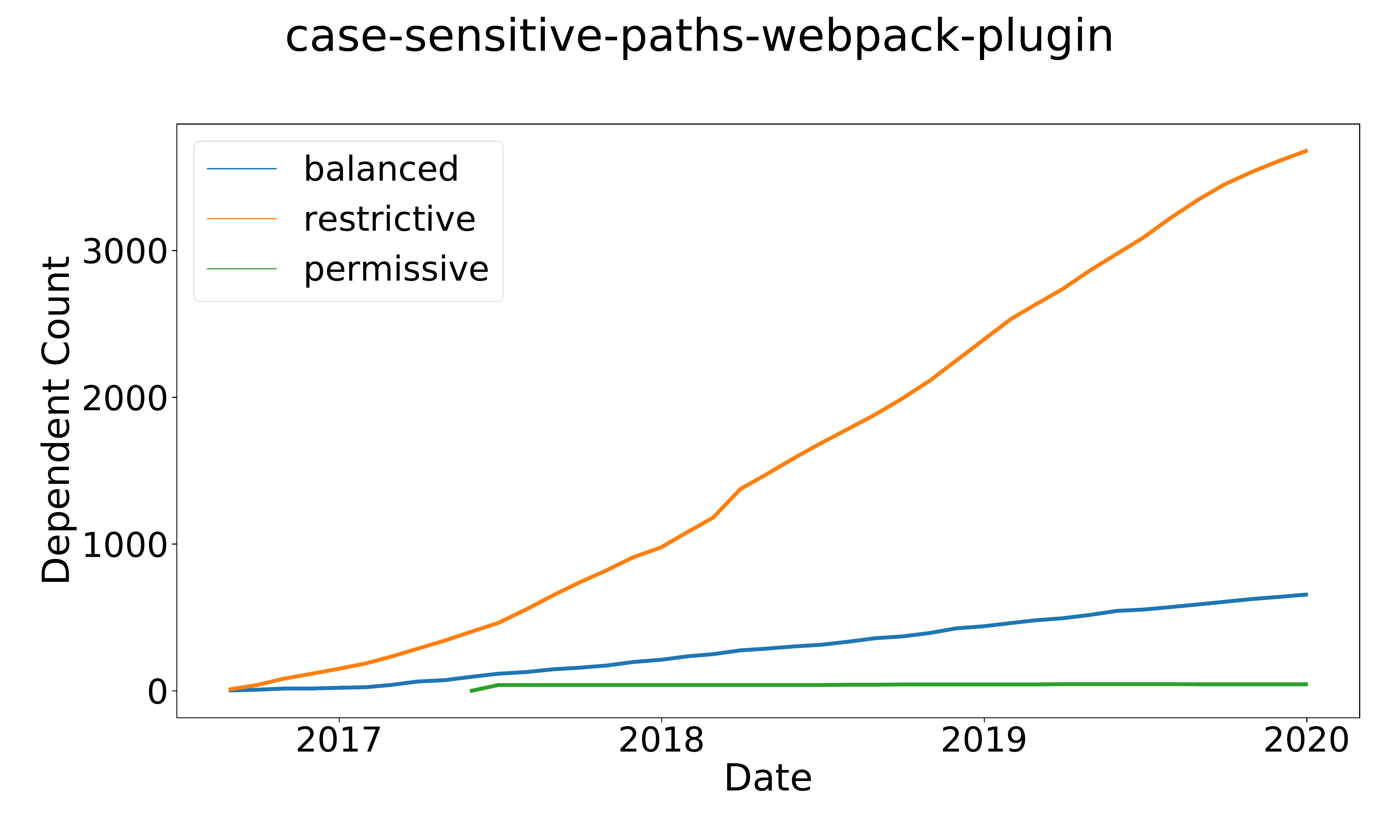}
	\end{subfigure}
	\hspace{-1mm}%
	\begin{subfigure}[b]{0.5\linewidth}
		\includegraphics[width=\linewidth]{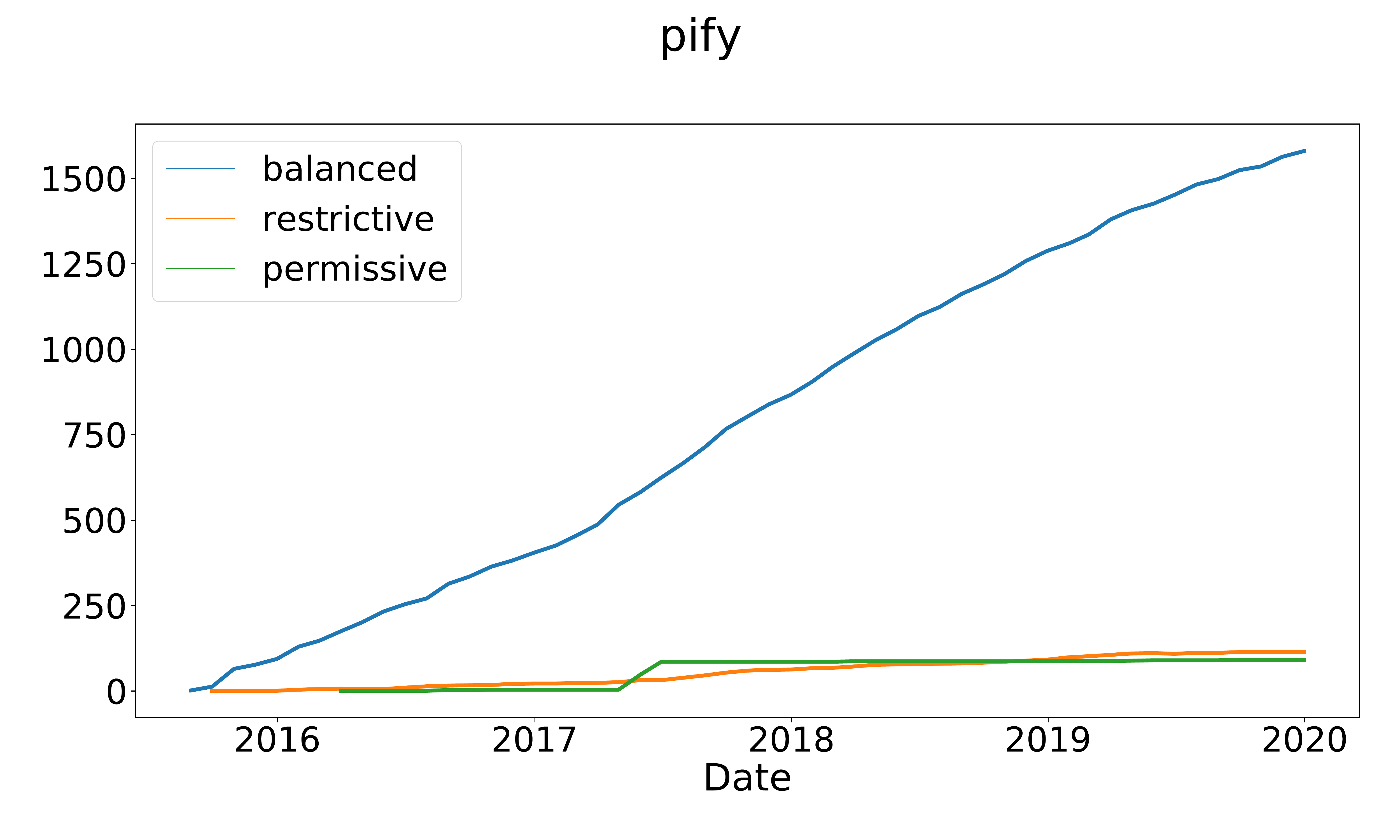}
		
	\end{subfigure}
  \caption{Example packages for which dependents follow the previously popular update strategy}
  \label{fig:previously_popular}
\end{figure}

\rev{One common evolution pattern is the tendency of dependents to follow the previously popular update strategy (i.e. agreement on the common update strategy does not change throughout the package's lifecycle). This evolution pattern was observed across the dependents of all package groups as shown in Figure~\ref{fig:previously_popular}.} We observed this pattern for 18 instances of balanced packages, 28 instances of permissive packages and 6 instances of restrictive packages. This finding aligns with the observation of Dietrich et al., which state that packages tend to stick to their dependency habits for a particular dependency \cite{dietrich2019dependency}. It is also worth noting that this behavior was observed in example packages specialized to all of the three update strategies, meaning it is not a result of dependents merely using the default npm update strategy (which leans toward the balanced update strategy). 

\conclusion{Finding \#1: For many npm packages, the common update strategy of its dependents remains consistent.}

\begin{figure}[tbh!]
	\begin{subfigure}[b]{0.5\linewidth}
		\includegraphics[width=\linewidth]{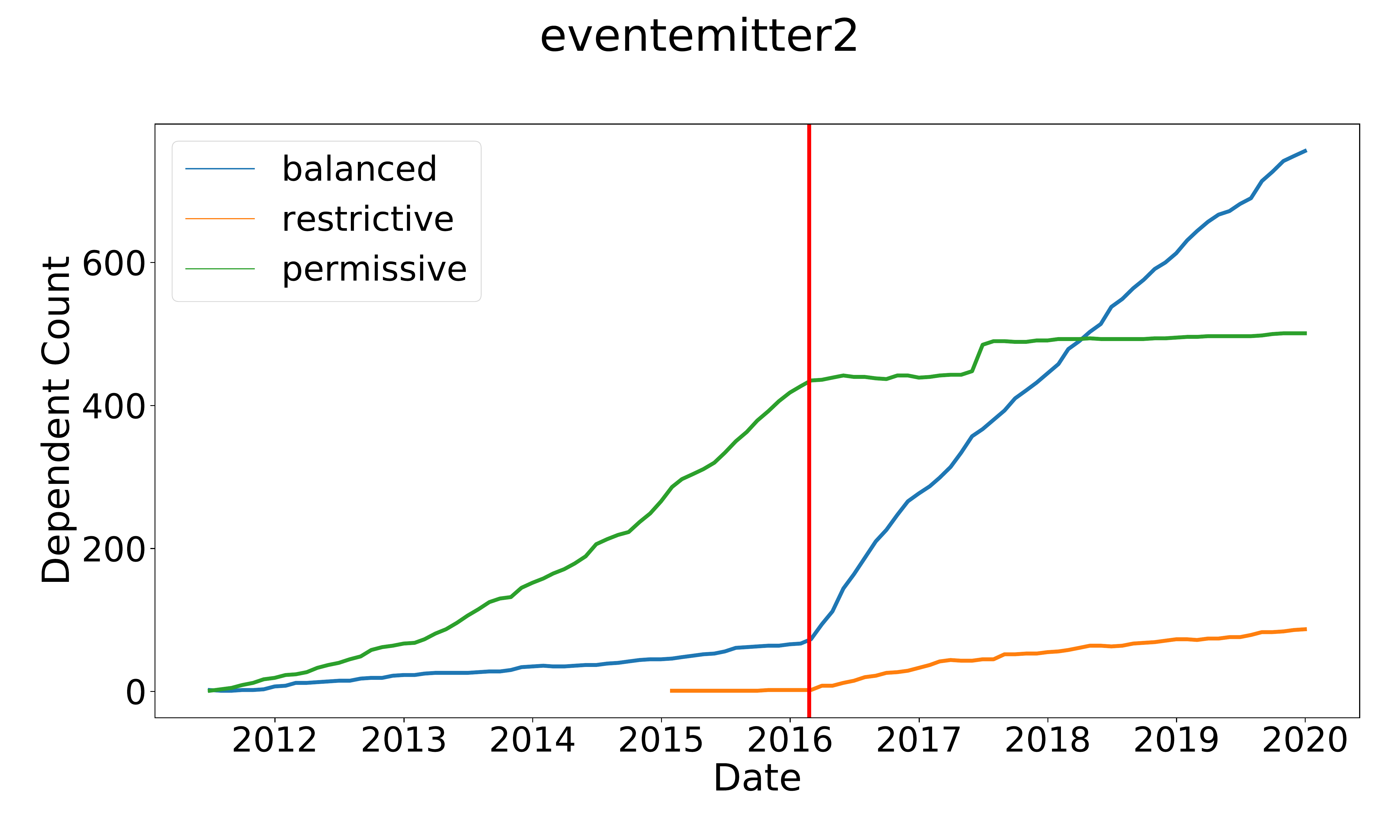}
	\end{subfigure}
	\hspace{-1mm}%
	\begin{subfigure}[b]{0.5\linewidth}
		\includegraphics[width=\linewidth]{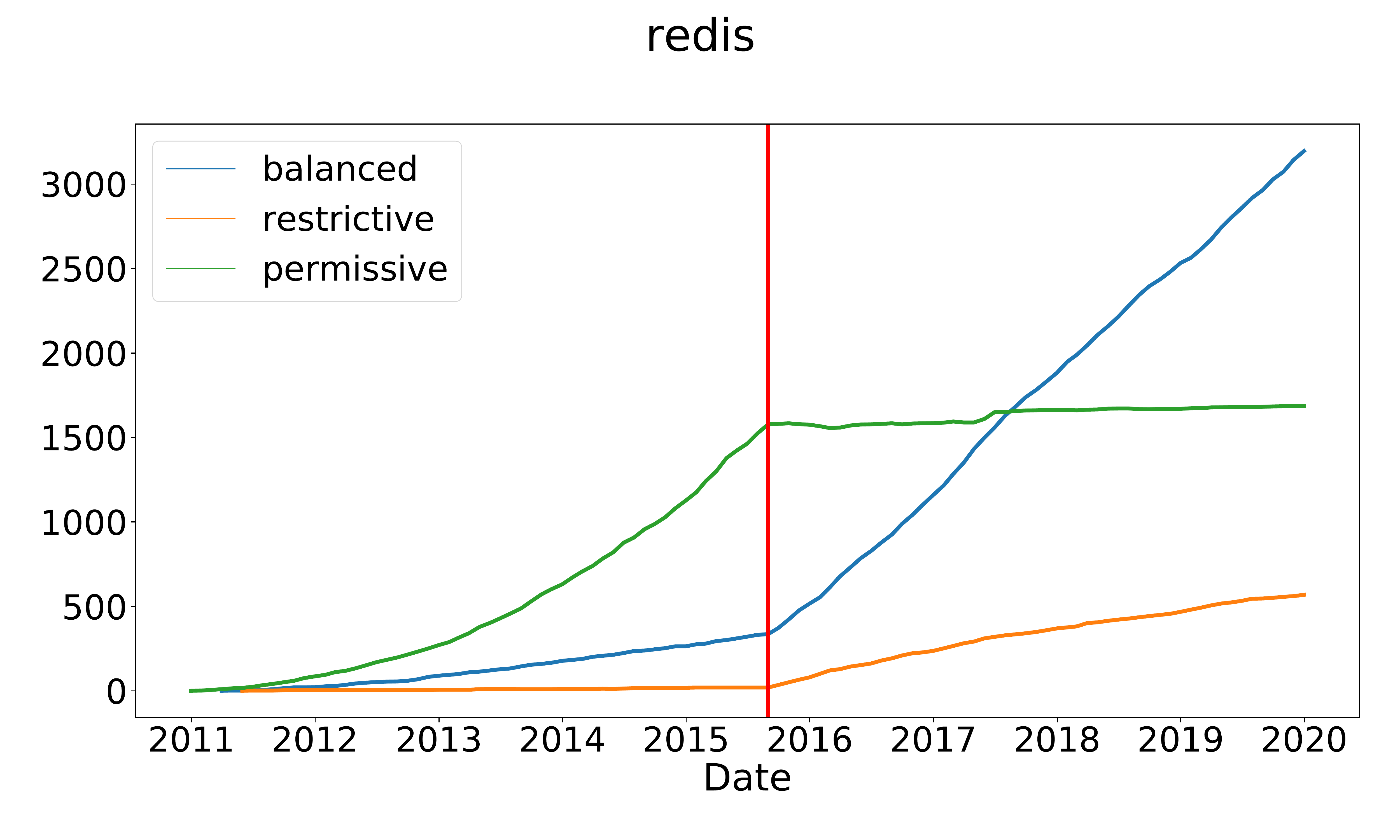}
	\end{subfigure}

  \caption{\rev{Example packages for which the dependent strategy shifts at the 1.0.0 release mark (red vertical line)}}
  \label{fig:strategy_shift}
\end{figure}

The pre-1.0.0 release versions of an npm package is considered to be unstable due to its initial development stage. \rev{However, Decan et al. studied package usage for pre-1.0.0 releases and found that there is no considerable difference between the number of dependents for pre-1.0.0 and post-1.0.0 releases \cite{decan2021lost}.} In our sample dataset, we observed an interesting phenomenon when a package releases its 1.0.0 version. When a highly used pre-1.0.0 package releases switches to a post-1.0.0 status, there is a very observable shift from permissive to balanced update strategies among its dependents. The examples in Figure~\ref{fig:strategy_shift} clearly show the impact of the 1.0.0 release (red line) on the update strategy evolution. While there are still dependents that use the permissive update strategies after the 1.0.0 release, the majority of new dependent relationships shift to the balanced strategy. The pattern generally appears when the pre-1.0.0 releases were already used by many dependents (which is why it can not be observed in the examples of Figure~\ref{fig:previously_popular}). \rev{This pattern may have occurred because the npm community is less accepting of the SemVer standard as it pertains to pre-1.0.0 releases and does not believe pre-1.0.0 dependencies should necessarily be pinned to a particular version \cite{decan2021lost}.} This particular pattern is observed for 12 instances of balanced packages, 6 instances of permissive packages and 15 instances of unspecialized packages.

\conclusion{Finding \#2: For highly used pre-1.0.0 packages, the release of the 1.0.0 version can change the common update strategy from permissive to balanced.}

\begin{figure}[tbh!]
	\begin{subfigure}[b]{0.5\linewidth}
		\includegraphics[width=\linewidth]{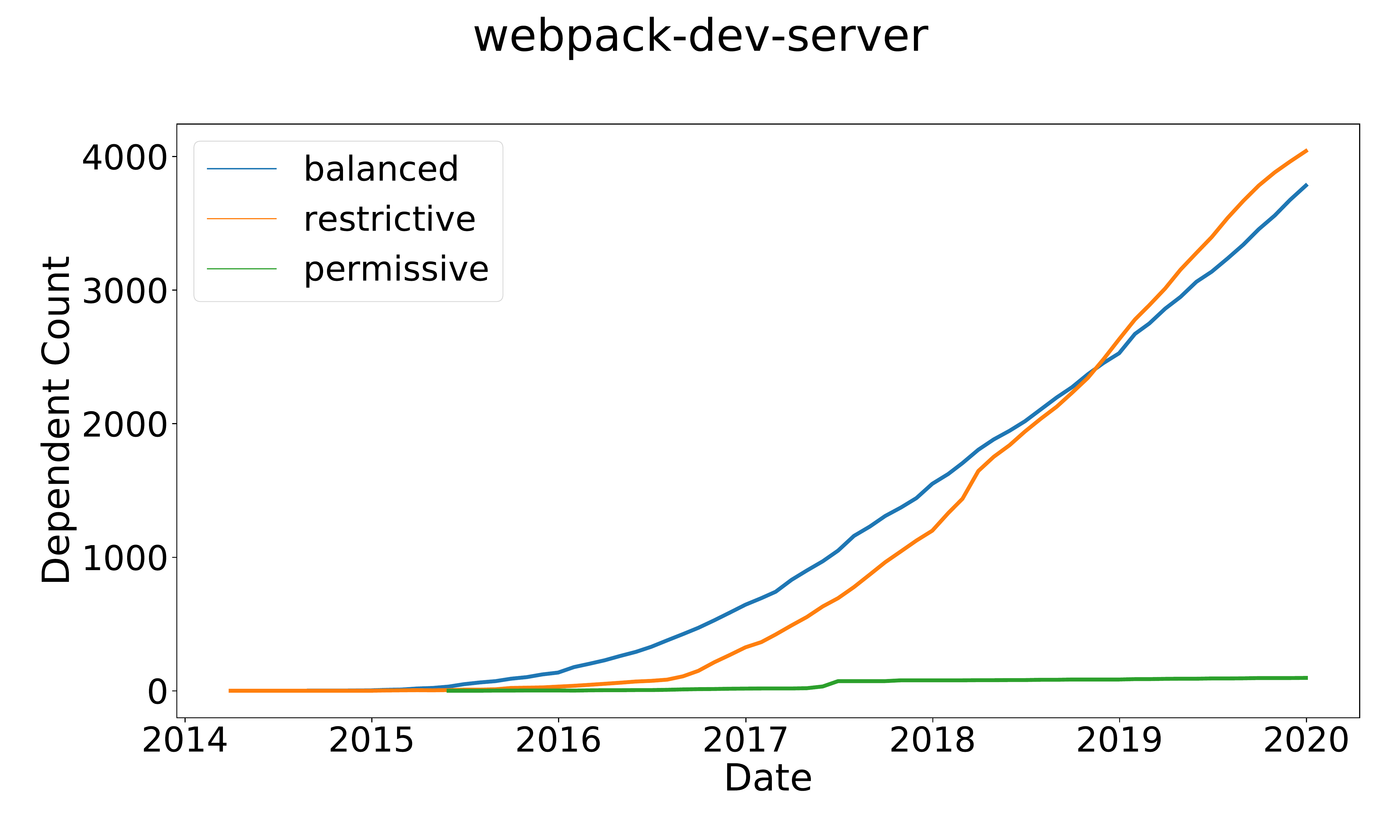}
	\end{subfigure}
	\hspace{-1mm}%
	\begin{subfigure}[b]{0.5\linewidth}
		\includegraphics[width=\linewidth]{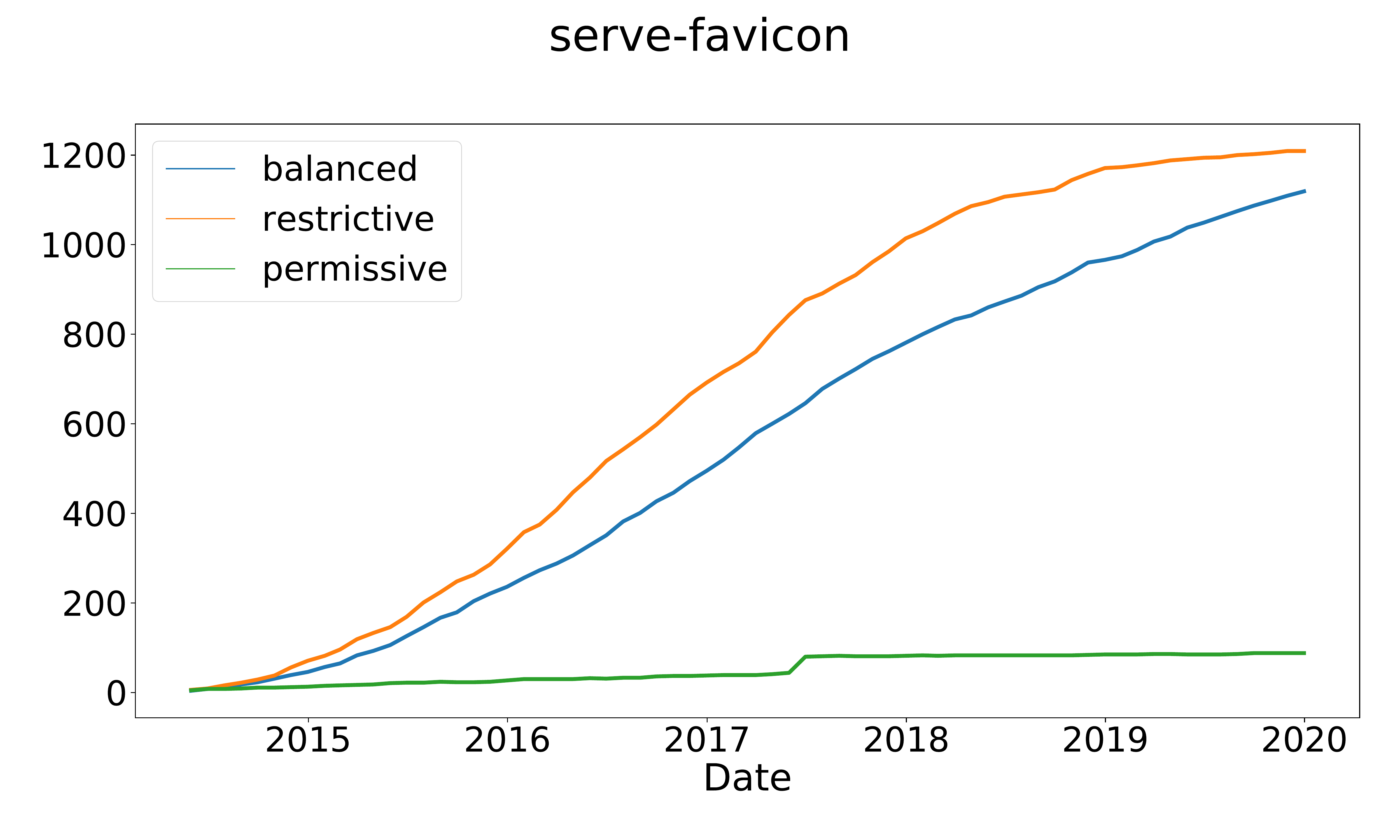}
		
	\end{subfigure}
  \caption{\rev{Example packages for which there is a weak agreement on the restrictive update strategy}}
  \label{fig:weak_agreement}
\end{figure}

The evolution of update strategies for dependents of packages specialized toward the restrictive update strategy exhibits unusual and anomalous behavior that is not observed in the other two package groups (balanced and permissive). First of all, it is more common to see packages that have a borderline agreement in the restrictive cases. The examples in Figure~\ref{fig:weak_agreement} show that while the evolution of update strategies for these packages ultimately leads the restrictive update strategy as the dominant one, a very considerable number of dependents still use the balanced update strategy when depending on these packages. Restrictive update strategies are a reluctant response to breaking changes or other problems with automatically updating to new minor versions of the dependency \cite{jafari2020dependency}. Therefore, the observed disagreement on the restrictive update strategy can happen because either a portion of the community is not aware of an existing issue with the package or because the issues do not equally affect all dependents. We observed this pattern in 10 instances of restrictive packages and 6 instances of unspecialized packages.

\conclusion{Finding \#3: Even when restrictive update strategies are the majority, they experience weaker agreements due to many dependents opting for balanced update strategies.}

\begin{figure}[tbh!]
	\begin{subfigure}[b]{0.5\linewidth}
		\includegraphics[width=\linewidth]{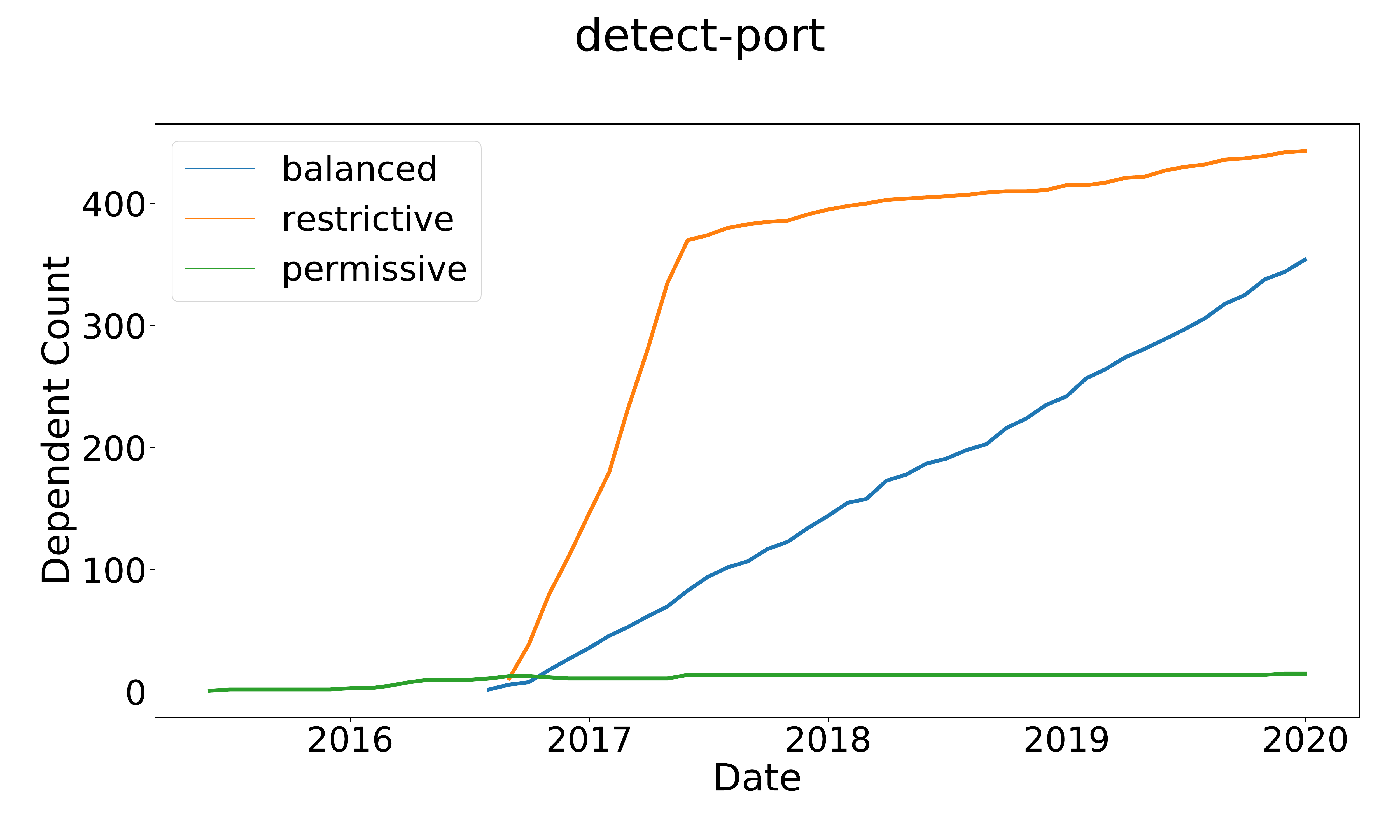}
	\end{subfigure}
	\hspace{-1mm}%
	\begin{subfigure}[b]{0.5\linewidth}
		\includegraphics[width=\linewidth]{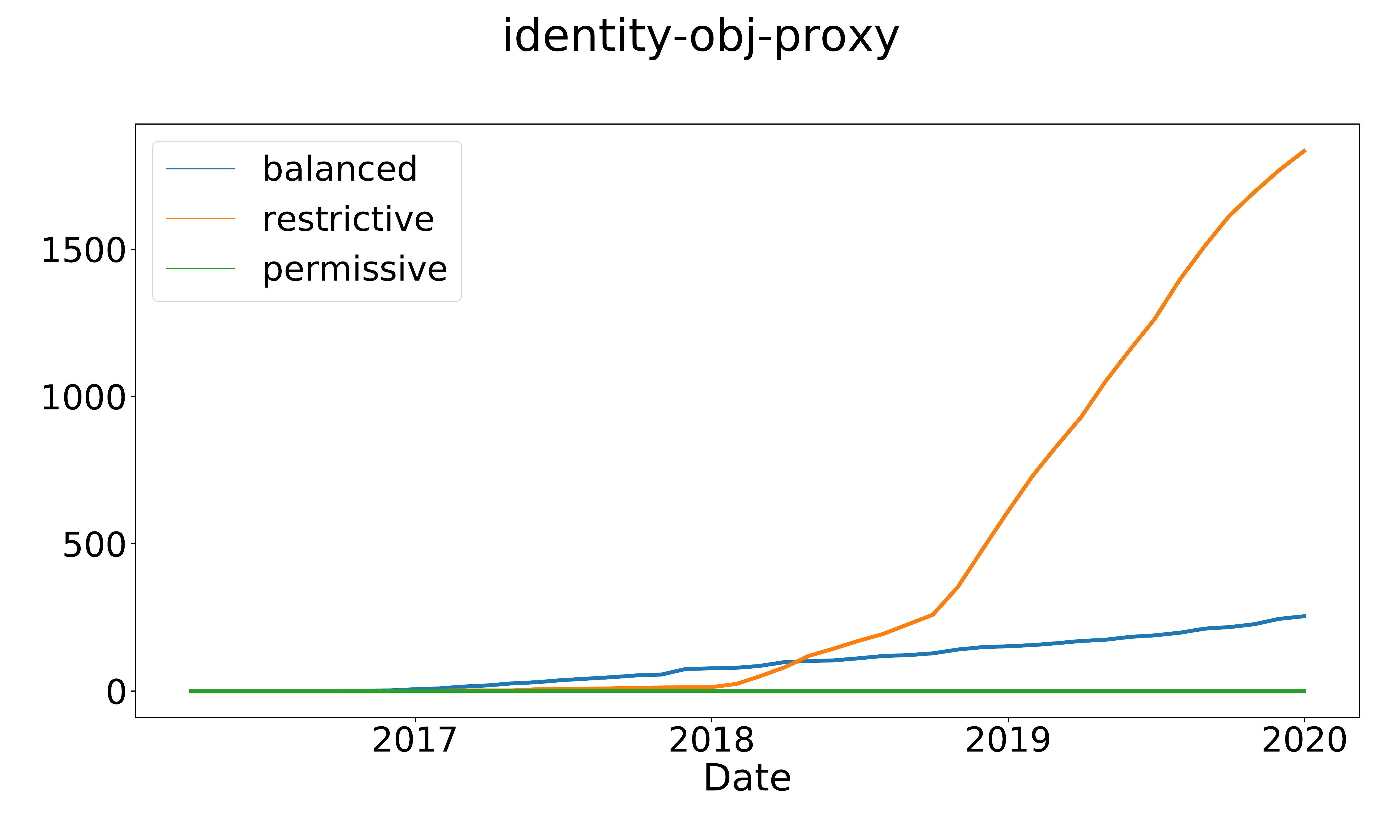}
	\end{subfigure}
   \begin{subfigure}[b]{0.5\linewidth}
		\includegraphics[width=\linewidth]{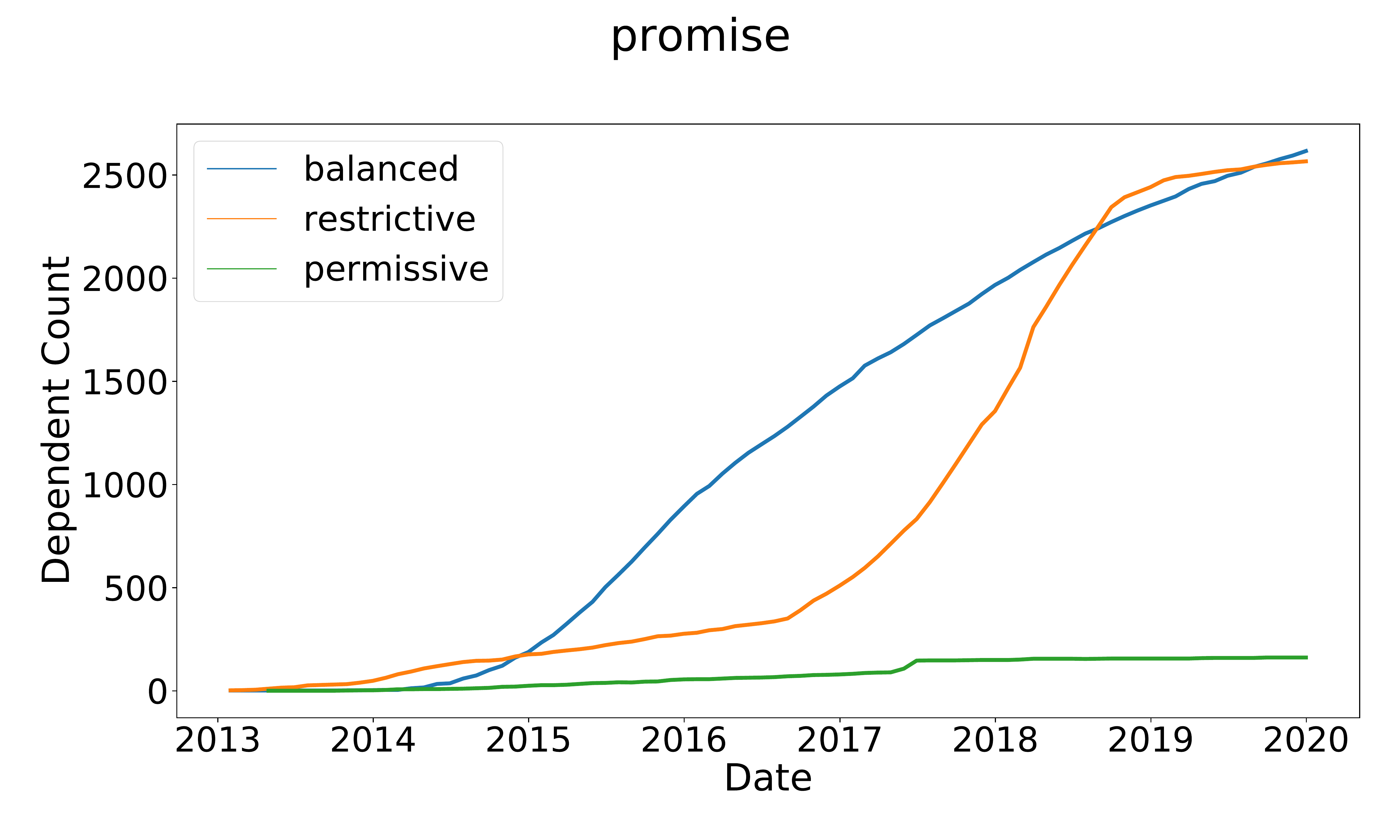}
	\end{subfigure}
	\hspace{-1mm}%
	\begin{subfigure}[b]{0.5\linewidth}
		\includegraphics[width=\linewidth]{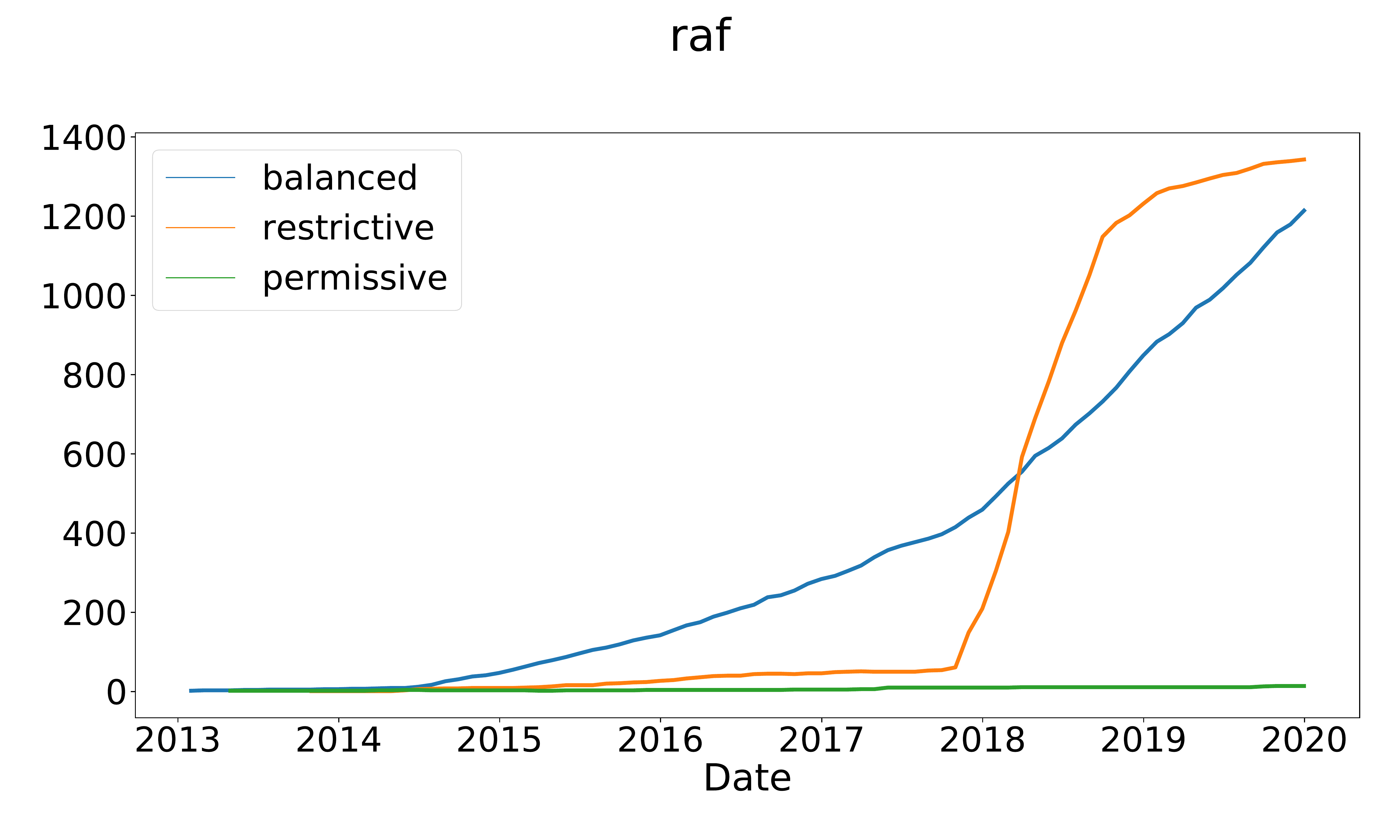}
		
	\end{subfigure}
  \caption{Example packages for which the restrictive update strategy exhibits anomalous behavior}
  \label{fig:anomalous}
\end{figure}

The other unusual observation for restrictive dependency update strategies is their anomalous evolutionary behavior. For example, in the evolution of update strategies for packages in Figure~\ref{fig:anomalous}, we see a sudden spike in the number of restrictive update strategies starting at a specific point in time that is very dissimilar to the gradual increase of the other two update strategies. This can happen if a particular event in time (perhaps a breaking change) causes a shift in community perception toward that package. The observation may also be due to \rev{a new set of} dependents with more conservative strategies that started using the package for the first time. The latter is more likely in cases such as \textit{detect-port} and \textit{identity-obj-proxy}. Alternatively, in cases such as \textit{promise} and \textit{raf} where the community moves back to the balanced strategy after a certain amount of time, the former explanation is more likely. We found such \rev{anomalous behavior} in 4 \rev{instances} of balanced packages, 8 instance of restrictive packages and 3 instances of unspecialized packages.

The findings for the evolution analysis of the restrictive update strategy warrants a closer look into the capability to identify them using package characteristics. While RQ1 presents the overall performance of our model, the per-class evaluation results can provide further insight. Table~\ref{tab:detailed_eval} presents the precision, recall and F1-score for each of the 3 main classes of the model, along with the unspecialized label (since some npm packages are not specialized toward any update strategy and they must also be included in the evaluation). We have also included the per-class F1-scores for the two baseline models for comparison. F1-Stratified denotes the F1-score for the stratified baseline and F1-Balanced denotes the F1-score for the Balanced only model. While our model outperforms the baseline for all 3 main classes, the restrictive class seems to be more difficult to predict across all models. Specifically, our model achieves high precision but low recall for the restrictive cases, indicating the model is mostly correct when classifying a restrictive package, but it also misses many of the other restrictive cases. The challenges in predicting the restrictive update strategy can be due to the limited number of packages specialized toward the restrictive strategy in the ecosystem (7\% of our main dataset) or due to the incidental nature of such strategies that are caused due to target package misbehavior (e.g. breaking changes) rather than its characteristics.

\begin{table*}
	\small
	\centering
		\caption{Per-Class Evaluation}
		\label{tab:detailed_eval}
		\begin{tabular}{l|r|r|r|r|r}
			\toprule
			Class Label & Precision & Recall & F1-score & F1-Stratified & F1-Balanced\\
			\midrule
			Balanced & 80\% & 84\% & 82\% & 54\% & 70\% \\
			Permissive & 74\% & 85\% & 79\% & 29\% &0\%\\
			Restrictive & 77\% & 32\% & 45\% & 6\% &0\%\\
			Unspecialized & 47\% & 33\% & 39\% & 9\% &0\%\\
			\bottomrule
		\end{tabular}
\end{table*}

Further examination of the anomalous behavior in the evolution of restrictive update strategies necessitates a qualitative approach. Thus, we manually analyze 1) The npm registry \cite{npmjs}, 2) The snyk open source advisory \cite{snykio} and 3) The GitHub repositories of the 40 sampled packages in the restrictive group. The npm registry provides information regarding installation notes, current weekly downloads of each version and build status badges. The snyk advisory provides information about known security vulnerabilities along with a package health score that considers security in addition to package popularity and maintenance. The GitHub repository provides the development history of the package. Using the repository information, we can filter created and resolved issues during a specific historical window to identify breaking changes that may correspond to the rise of a restrictive update strategy for that package.

We started with the npm registry page of each package to search for mentions of SemVer non-compliance from maintainers of the package. We hypothesized that one reason for the popularity of restrictive update strategies for this group of packages would be the official statements by package maintainers that indicate their misalignment with SemVer compliance. None of the 40 packages had stated anything about the recommended update strategy. Thus, we can speculate that the choice of a restrictive update strategy is solely on the dependents' side. One interesting observation was the maintainer's recommendation to install their packages as a development dependency, as opposed to a runtime dependency, in \rev{65\% of these packages}. Since our dataset is filtered to only include runtime dependency relations, many dependents have obviously not followed this recommendation.

The snyk advisory provides a package health score that combines security, popularity, maintenance and community factors into a single metric \cite{snykio}. More importantly, snyk is a vulnerability dataset that catalogs low, medium, high and critical severity vulnerabilities recorded for each version of a package. We hypothesized that vulnerable releases will encourage package dependents to restrict their update strategies while they wait for a fix to be released. With the exception of the ``webpack-dev-server'' package in which 144 versions were infected by a high severity vulnerability, the rest of the packages in our sample had no recorded vulnerabilities. In simpler terms, we could not find sufficient evidence that indicate restrictive update strategies are mainly the result of vulnerable releases. 

The GitHub repository of the packages allows open access to the development history of the package, along with recorded issues and feature requests. 

We hypothesized that breaking changes from new releases may be a reason why dependents opt for a more restrictive update strategy. To this aim, we searched through repository issues created for each package during the one year window in which we observed a rise in restrictive update strategies from the dependents of that package. We found concrete evidence of breaking updates in 18 of the 40 packages in the restrictive group. Not all breaking updates lead to newly created issues about the problem, so our findings are actually a lower bound on the number of packages that experience breaking changes. In fact, out of the 22 packages with no evidence of breaking changes, 11 packages had low activity (less than 50 open and closed issues combined) or no activity in their repository issue tracker throughout the project's history. Table~\ref{tab:detailed_eval} presents example issues from package repositories where users voice their concerns about breaking changes (or other problems) caused by updating to a new version. These findings align with prior research that identifies breaking changes and dependency misbehavior as highly influential factors in restrictive dependency update policies by the dependents \cite{jafari2020dependency}.

\begin{table*}
	\small
	\centering
		\caption{Examples of created issues that correspond with a rise of restrictive update strategies}
		\label{tab:examples_breaking}
		\begin{tabular}{l|l|l}
			\toprule
			Package Name & Issue Date & Issue Title\\
			\midrule
			postcss-loader & Jan 2017 & ``v1.2.1 runs fine, but v1.2.2 throws error'' \\
			eslint-plugin-jsx-a11y & Jun 2016 & ``Exception after update to 1.4.0'' \\
			jest-resolve & Dec 2018 & ``medium severity vulnerability […] introduced via jest @23.6.0'' \\
			eslint-loader & Apr 2015 & ``npm error after update to version 0.11.0'' \\
			fsevents & Feb 2017 & ``breaking change in 1.1.0'' \\
			\bottomrule
		\end{tabular}
\end{table*}

\conclusion{\rev{Finding \#4: Restrictive update strategies exhibit a more erratic evolutionary behavior that corresponds to breaking changes, making them harder to predict}}

\section{Implications}
\label{sec:Implications}
We present actionable implications for both practitioners (developers and package maintainers) and researchers in the field.

\subsection{Implications for Practitioners:}
The package characteristics model presented in this study has been shown to outperform the default balanced update strategy in npm (RQ1). \textbf{The \rev{predictions of the }model can be used as a recommendation for developers \rev{to help them in} deciding on a suitable dependency update strategy for a package. Alternatively, practitioners \rev{can rely on} the most important features such as release status, dependent count and age (RQ2) to \rev{aid their dependency update strategy selection.}} For example, using packages with a smaller number of dependents poses an inherent risk of not yet having an agreed upon update strategy in the community. In addition to the number of dependents, the prominence of those dependents should also be taken into account.

The release status of a package (pre-1.0.0 vs. post-1.0.0) has shown to be a relevant feature in identifying the common update strategy (RQ2) and there is an observable shift from the permissive update strategy to the balanced strategy when the 1.0.0 version is released (RQ3). The use of permissive constraints for pre-1.0.0 packages shows that developers in the npm community do not fully align with the SemVer standard for pre-1.0.0 releases. It is also a testament to the relatively high popularity of some pre-1.0.0 packages. We looked at the number of dependents for both the pre-1.0.0 and post-1.0.0 packages and found that while post-1.0.0 packages have a median of 4 dependents, pre-1.0.0 have a median of 3 dependents. This is surprising as SemVer considers pre-1.0.0 initial development releases to be unstable by nature and depending on them poses an inherent risk. Yet, a considerable portion of developers are already using such packages as dependencies. This confirms the findings of \rev{Decan et al. \cite{decan2021lost}} and highlights the importance of initial development releases for package maintainers. \textbf{Package maintainers should assume that initial development releases may already be used by dependents which could be stakeholders in future changes.}

While studying the evolution of dependency update strategies, we observed many instance where the initially established update strategy was also selected by new dependents, creating a compounding effect that ultimately leads to a clearly dominant dependency update strategy for dependents of that package. We did not find significant evidence of target packages recommending a particular update strategy to their users and this continuous trend was observed for all 3 types of update strategies (i.e. it can not simply be attributed to the use of the default balanced update strategy). Therefore, this behavior likely stems from independent decisions from package dependents, some of which may consider the previously common update strategy to be the best one. \textbf{Ecosystem maintainers should be attentive to the early adopter community of their packages as the first impressions set by the initial community can have long-lasting influence on how new dependents use their package.}

\subsection{Implications for Researchers:}

\rev{While the package characteristics model in this study can be leveraged to predict the suitable dependency update strategy (RQ1), there are other characteristics to explore. Further research is needed to extract and \textbf{look into other features such as the package downloads count, code complexity, the experience level of package maintainers and the quality of the documentation to see if and how these features can improve the model}.} Additionally, since we know that restrictive update strategies may be influenced by specific events rather than package characteristics (RQ3), future work is needed to cross-reference the time of the change with relevant events in the repository such as a bug/vulnerability fix or a newly opened issue to understand how such events can influence a change in the dependency update strategy. \textbf{We should also look at the frequency of change and the duration between changes in the dependency update strategy to better understand whether some events such as breaking changes have long-term impact on the trust of a particular package}.

\rev{The current model proposes a predicted update strategy based on the characteristics of a target package. However, it is beneficial to know the confidence in the recommended update strategy and the rankings of the non-recommended alternatives. While developers can use the important features discovered in this study as the basis for their own judgment, \textbf{a probabilistic model that complements the predictions by presenting a ranking of recommended update strategies can prove useful.}}

Not knowing why different dependency update strategies occur in a package creates data noise when analyzing the strategies. We previously \rev{discussed} how npm default constraints for newly added dependencies (RQ2) \rev{create} a challenge when analyzing the wisdom of the crowds since we do not fully know whether the developer chose the constraint or simply trusted the default update strategy. Using the balanced strategy can be traced back to meticulous planning by the dependent or a simple disregard toward dependency maintenance. \textbf{A valuable avenue for research is to study how much the ecosystem is impacted by developer decisions versus ecosystem policies, such as default dependency constraints.}

\rev{Restrictive update strategies are a response to issues such as breaking changes when updating dependencies. However, the entire dependent community of a package may not be equally aware or equally affected by such issues, which leads to weaker agreements on the restrictive update strategy (RQ3). In the wisdom of the crowds model, a high level of restrictive strategies (and their underlying cause) may be disregarded simply because they do not represent the majority. \textbf{An improved version of the model presented in this study can allow the specialization threshold to differ per each class to allow a strategy-sensitive model that is tuned to better predict the probability of a particular update strategy.}}

\section{Related Work}
\label{sec:Related Work}
To the best of our knowledge, there is no other work that utilizes package characteristics to predict the most suitable dependency update strategy and studies the impact of those characteristics on the selected strategy. The related work for our study is comprised of research that focuses on dependency update strategies, studies that focus on relevant characteristics in selecting dependencies and research in the npm ecosystem supply chain.
\\

\noindent\textbf{Dependency update strategies:}

Decan and Mens conducted an empirical study to compare SemVer compliance across four software ecosystems including npm \cite{decan2019package}. They proposed an update strategy based on ``the wisdom of the crowds'' to help developers choose the best dependency update strategy. They accomplished this by analyzing the dependency constraints of all dependents of a package and recommending the most common update strategy. This study is the most relevant to our work as it uses past dependency decisions to predict the most common update strategy in the future. However, the work of Decan et al. does not use package characteristics for prediction and requires a complete and updated dependency graph of the npm ecosystem, making it unscalable in practice. Our method is scalable as it only looks at the current characteristics of the package and does not need dependency information from the dependents. More importantly, our work is the first to study the relationship between package characteristics and the predicted dependency update strategy. \rev{In another study, Decan et al. empirically investigated the pre-1.0.0 versions and their usage in 4 software ecosystems. They found that there is no practical difference between the usage of pre-1.0.0 and post-1.0.0 versions but ecosystems are more permissive than SemVer guidelines when it comes to using pre-1.0.0 versions \cite{decan2021lost}.}

Dietrich et al. studied dependency versioning practices across 17 software ecosystems including npm \cite{dietrich2019dependency}. Their study is complemented by a survey of 170 developers. They found that most ecosystems support flexible versioning practices but developers still struggle to manage the trade-offs between the predictability of more restrictive update strategies and the agility of more flexible ones. Feedback from more experienced developers suggest they favor the stability that accompanies restrictive update strategies. Dietrich et al. did not look at how package characteristics can impact the selected dependency update strategy and how such package characteristics can be used to guide developers towards the suitable strategy.

Jafari et al. empirically studied problematic dependency update strategies in JavaScript projects \cite{jafari2020dependency}. They cataloged and analyzed 7 dependency smells including restrictive constraints and permissive constraints. Their findings indicate that while smells are prevalent, they are localized to a minority of each project's dependencies. Through a developer survey, they highlighted the negative impacts of such update strategies and they also quantified the reasons for their existence. They found that such alternative update strategies are often the result of dependency misbehaviour or issues in the npm ecosystem. While Jafari et al. did not look at the impact of package characteristics on dependency update strategies, their work highlights the importance of studying such characteristics to understand why some npm packages implicitly push their dependents to use non-balanced dependency update strategies.
\\

\noindent\textbf{Package characteristics for selecting dependencies:}

Bogart et al. performed an empirical study on three software ecosystem including npm to study how developers make decisions in regard to change and change-related practices \cite{bogart2016break}. In their interview with 28 developers, they found that various signals are used to select dependencies. These include the level of trust on the developers of the package, activity level, user base, project history and artifacts such as documentation. The respondents believed such characteristics to be important in deciding what package to depend on, but the study did not look at how package characteristics can influence the chosen dependency update strategy.

Vargas et al. surveyed 115 developers to study the factors that impact the selection of dependency libraries \cite{larios2020selecting}. They observed several technical factors such active maintenance, code stability, release frequency, usability and performance to be relevant factors. The authors also observed human factors such as community perception and popularity along with economic factors such as license and cost of ownership to be contributing factors in selecting a dependency.

Pashchenko et al. interviewed 25 industry practitioners to investigate the influence of functional and security concerns on decision making with regards to software dependencies \cite{pashchenko2020qualitative}. The authors found that developers rely on high-level information that demonstrates the community support of a library such as popularity, commit frequency and project contributors. Developers prefer libraries that are safe to use and do not add too many transitive dependencies. The authors observed that dependency selection is often assigned to more skilled members of the team.

Haenni et al. conducted a survey and asked developers about their information needs with respect to their upstream and downstream packages \cite{haenni2013categorizing}. Developers stated that they consider factors such as popularity, documentation, license type, update frequency and compatibility when looking for a new dependency. The authors also found that in practice, developers monitor news feeds, search through package websites and blogs and run their unit tests to achieve these goals.

The four aforementioned studies all focus on relevant characteristics in selecting a package as a dependency. They do not study the impact of these characteristics on the update strategy used for each dependency.
\\

\noindent\textbf{The npm ecosystem supply chain:}

Zimmerman et al. studied how the packages and package maintainers in npm have the potential to impact large chunks of the ecosystem \cite{zimmermann2019small}. They looked at a collection of more than five million package versions in npm and observed that installing an average npm package is the equivalent of implicitly trusting 79 packages and 39 maintainers. Additionally, they realized that up to 40\% of npm packages depend on a vulnerable package with a publicly disclosed vulnerability. The authors found that, among other things, locking dependencies exacerbates the security issues in the ecosystem since it hinders the automatic adoption of a vulnerability fix. 

Zerouali et al. empirically analyzed the technical lag in the npm ecosystem and its relationship to dependency update strategies \cite{zerouali2018empirical}. The authors used a subset of the libraries.io dataset comprised of 610K packages and over 4.2 million package versions. They found that while npm packages are frequently updated, dependencies are rarely added or removed. They also discovered that restrictive dependency update strategies are the main culprit for technical lag in the ecosystem.

Cogo et al. conducted an empirical study on same-day releases in the npm ecosystem \cite{cogo2021empirical}. They found same day releases to be common in popular packages, interrupting a median of 22\% of regular release schedules. More importantly, they observed that 32\% of such releases encompass even larger changes than their prior (planned) release. In general, downstream dependents of popular packages tend to automatically adopt same-day releases due to their dependency update strategies. The authors believe same-day release to be a significant occurrence in the npm ecosystem and dependency management tools should consider flagging such releases for downstream dependents.

Chowdhury et al. studied trivial packages in the npm ecosystem (micro-packages with only a few lines of code) \cite{chowdhury2021untriviality}. They found that close to 17\% of the packages in the ecosystem can be considered trivial, but removing one of these packages can impact up to 29\% of the entire ecosystem. While such small packages are small in size and complexity, they are responsible for a high percentage of API calls. Trivial packages play an important and significant role in the npm ecosystem.

\section{Threats to Validity}
\label{sec:Threats to Validity}
This section discusses the threats to the validity of our study.

\noindent\textbf{Threats to construct validity} consider the relationship between theory and observation, in case the measured variables do not measure the actual factors. Our specification of dependency update strategies considers version constraints and assumes developers use the official npm registry to fetch their dependencies. In reality, developers can look outward and use external sources to fetch dependencies (e.g. direct link to Github repository). One issue with such cases is that the update strategy could change depending on the contents of the external source. For example, linking to the master branch is equivalent to a permissive update strategy and linking to a specific release is equivalent to a restrictive update strategy. Another issue is that there is no way to identify all package dependents if the package is hosted on an external link. In order to study both the dependencies and the dependents of the packages, our study only considers packages hosted on the official npm registry and dependencies pointing to other packages in the npm ecosystem. Additionally, we assume the information provided by the libraries.io dataset \cite{libraries.io2020} is accurate, and this assumption has been verified by other researchers \cite{decan2019empirical}.

\noindent\textbf{Threats to internal validity}
refer to internal concerns such as experimenter bias and error. The npm ecosystem is very large and susceptible to noisy/toy packages. We disregard packages with less than 2 dependents which removes unused packages from our dataset. We also manually remove multiple spam packages (and their dependencies) which had the sole purpose of depending on every other package in the ecosystem (Section~\ref{sec:Methodology and Data}). In order to train our model, we use 19 features that we believe to influence dependency decisions based on the literature. In reality, there may be other relevant information for deciding on the dependency update strategy that were not captured (or not feasible) using our feature set. For example, developers can change dependency update strategies following a recommendation from a senior member of the team or because the specific section of the code relying on the dependency is critically important. We believe our features to be suitable since we cross-referenced the relevant characteristics for dependency selection and management that we found in the literature, with the package characteristics available in the npm registry and the code repository. \rev{We discovered features with missing data in the repository fields of the libraries.io dataset, warranting a look into the accuracy of the dataset. For many features (e.g. Dependency Count) the null value was used to denote zero as the minimum value starts at one. However, in 3 out of the 19 features selected for our model (Repository Stars Count, Repository Size, and Repository Open Issues Count), we found missing values where a value of zero was also present. We took a sample of 1000 packages that had missing data corresponding to the three features and realized 96.1\% of these packages do not have a working repository link (repository no longer exists). Section~\ref{sec:Methodology and Data} explains how we handled missing values in our dataset. Our findings regarding the accuracy of the libraries.io dataset corroborates the previous analysis of Decan et al. in which they manually cross-checked the libraries.io dataset against their own collected metadata from the npm registry and verified its accuracy \cite{decan2019empirical}.}

\noindent\textbf{Threats to external validity}
concern the generalization of our findings. The observed findings are specific to the npm ecosystem since previous research has shown that different ecosystems have different practices and cultural values \cite{bogart2016break,values2020}. However, the package characteristics, the methodology to extract the features and the update strategy to train the model can be replicated on other ecosystems that provide similar dependency information. In fact, since the libraries.io dataset \cite{libraries.io2020} used in this study utilizes the same schema to store metadata for other ecosystems such as PyPI and Maven, our replication package \cite{replication} can easily be used to replicate the study on other ecosystems. \rev{Additionally, the libraries.io dataset used in this study does not contain npm package data after January 2020. However, re-collecting the dataset for an entire ecosystem such as npm does not only require a lot of effort, but it is error-prone. The accuracy of the libraries.io dataset has previously been verified in the literature \citep{decan2019empirical}. More importantly, our study is more focused on the dynamics of dependency management in the npm ecosystem, rather than predicting the update strategy for the latest available version. Therefore, we believe the dataset to be suitable for our study. The findings of RQ3 are derived from a sample of 160 packages. While these packages are selected at random, we want to focus on packages with adequate historical dependent data. Therefore, our selection criteria requires packages to have more than 100 dependents, which threatens the generalizability of the results of this particular RQ to packages with a small number of dependents. As previously mentioned, the sample of 160 packages is not meant as a representative sample of the entire ecosystem. It is a convenience sample of highly used packages for an in-depth mixed-method study that is otherwise infeasible for such a large ecosystem.}

\section{Conclusion}
\label{sec:Conclusion}
In our study, we use a curated dataset of over 112,000 npm packages to collect and derive \rev{19} package characteristics from the their npm registry and code repository. We use these characteristics to train a model to predict the most commonly used dependency update strategy for each package. Based on the wisdom of the crowds principle, we believe the update strategy used by the majority to be favorable to the alternatives. We show that these characteristics can in fact be used to predict dependency update strategies. We analyze the most important features that influence the predicted update strategy and show how a change in these features influences the predictions. Developers should take note of the highly important characteristics and their impact when making dependency decisions about a package. The results show that our model outperforms the alternative of merely using the balanced update strategy in all instances. \rev{We complement the work with a manual analysis of 160 packages to investigate the evolutionary behavior of dependency update strategies and understand how they are impacted by events such as the 1.0.0 release or breaking changes.}


\bibliographystyle{ACM-Reference-Format}
\bibliography{main-base}

\appendix

\end{document}